\documentclass{aastex631}
\shorttitle{Determining the Magnetic Field of a Solar Active Region}
\shortauthors{Ishikawa et al.}
\graphicspath{{./}{figures/}}

\begin{document}

\title{Determining the Magnetic Field in the Atmosphere of a Solar Active Region
Observed by the CLASP2.1 Sounding Rocket Experiment}

\correspondingauthor{Ryohko Ishikawa}
\email{ryoko.ishikawa@nao.ac.jp}

\author[0000-0001-8830-0769]{Ryohko Ishikawa}
\affiliation{National Astronomical Observatory of Japan, \\
2-21-1 Osawa, Mitaka, Tokyo 181-8588 Japan}

\author[0000-0001-5131-4139]{Javier Trujillo Bueno}
\affiliation{Instituto de Astrof\'{i}sica de Canarias, E-38205 La Laguna, Tenerife, Spain}
\affiliation{Departamento de Astrof\'{i}sica, Universidad de La Laguna, E-38206 La Laguna, Tenerife, Spain}
\affiliation{Consejo Superior de Investigaciones Cient\'{i}ficas, Spain}

\author[0000-0002-9921-7757]{David E. McKenzie}
\affiliation{NASA Marshall Space Flight Center, Huntsville, AL 35812, USA}

\author[0000-0003-3034-8406]{Donguk Song}
\affiliation{Korea Astronomy and Space Science Institute 776, \\
Daedeokdae-ro, Yuseong-gu, Daejeon 305-348, Republic of Korea}
\affiliation{National Astronomical Observatory of Japan, \\
2-21-1 Osawa, Mitaka, Tokyo 181-8588 Japan}

\author[0000-0003-1465-5692]{Tanaus\'{u} del Pino Alem\'{a}n}
\affiliation{Instituto de Astrof\'{i}sica de Canarias, E-38205 La Laguna, Tenerife, Spain}
\affiliation{Departamento de Astrof\'{i}sica, Universidad de La Laguna, E-38206 La Laguna, Tenerife, Spain}

\author[0000-0001-9095-9685]{Ernest Alsina Ballester}
\affiliation{Instituto de Astrof\'{i}sica de Canarias, E-38205 La Laguna, Tenerife, Spain}
\affiliation{Departamento de Astrof\'{i}sica, Universidad de La Laguna, E-38206 La Laguna, Tenerife, Spain}

\author[0000-0002-8775-0132]{Luca Belluzzi}
\affiliation{Istituto ricerche solari Aldo e Cele Dacc\`{o} (IRSOL), Faculty of Informatics, 
Universit\`{a} della Svizzera italiana (USI), \\
CH-6605 Locarno, Switzerland}
\affiliation{Euler Institute, Universit\`{a} della Svizzera italiana (USI), CH-6900 Lugano, Switzerland}

\author[0000-0001-5612-4457]{Hao Li}
\affiliation{State Key Laboratory of Solar Activity and Space Weather, National Space Science Center, Chinese Academy of Sciences, Beijing, 100190, People's Republic of China}

\author[0000-0003-0972-7022]{Fr\'ed\'eric Auch\`ere}
\affiliation{Institut d'Astrophysique Spatiale, 91405 Orsay Cedex, France}

\author[0000-0001-9076-6461]{Christian Bethge}
\affiliation{Cooperative Institute for Research in Environmental Sciences, University of Colorado Boulder, Boulder, CO 80305, USA}

\author[0000-0002-8370-952X]{Bart De Pontieu}
\affiliation{Lockheed Martin Solar \& Astrophysics Laboratory, Palo Alto, CA 94304, USA}

\author[0000-0002-2093-085X]{Ryouhei  Kano}
\affiliation{National Astronomical Observatory of Japan, \\
2-21-1 Osawa, Mitaka, Tokyo 181-8588 Japan}

\author[0000-0003-1057-7113]{Ken Kobayashi}
\affiliation{NASA Marshall Space Flight Center, Huntsville, AL 35812, USA}

\author[0000-0002-4691-1729]{Adam R. Kobelski}
\affiliation{NASA Marshall Space Flight Center, Huntsville, AL 35812, USA}

\author[0000-0003-3765-1774]{Takenori J. Okamoto}
\affiliation{National Astronomical Observatory of Japan, \\
2-21-1 Osawa, Mitaka, Tokyo 181-8588 Japan}

\author[0000-0002-3770-009X]{Laurel A. Rachmeler}
\affiliation{National Oceanic and Atmospheric Administration, \\ 
National Centers for Environmental Information, Boulder, CO 80305, USA}

\author[0000-0003-2991-4159]{Taro Sakao}
\affiliation{Institute of Space and Astronautical Science, Japan Aerospace Exploration Agency, \\
Sagamihara, Kanagawa 252-5210, Japan}

\author[0000-0002-8292-2636]{Ji\v{r}\'i  \v{S}t\v{e}p\'an}
\affiliation{Astronomical Institute, Academy of Sciences of the Czech Republic,
25165 Ondrejov, Czech Republic}

\author[0000-0002-7219-1526]{Genevieve D. Vigil}
\affiliation{NASA Marshall Space Flight Center, Huntsville, AL 35812, USA}

\author[0000-0002-5608-531X]{Amy Winebarger}
\affiliation{NASA Marshall Space Flight Center, Huntsville, AL 35812, USA}



\begin{abstract}
We determine magnetic fields from the photosphere to the upper chromosphere combining data from the Hinode satellite and the CLASP2.1 sounding rocket experiment. 
CLASP2.1 provided polarization profiles of the Mg~{\sc ii} $h$ and $k$ lines, as well as of the Mn~{\sc i} lines around 2800~{\AA}, across various magnetic structures in an active region, containing a plage, a pore, and the edges of a sunspot penumbra.
By applying the Weak-Field Approximation (WFA) to the circular polarization profiles of these spectral lines, we obtain a longitudinal magnetic field map at three different heights in the chromosphere (lower, middle, and upper). 
This is complemented by data from Hinode (photospheric magnetic field), IRIS, and SDO (high-spatial-resolution observations of the chromosphere and corona).
We quantify the height expansion of the plage magnetic fields and find that the magnetic fields expand significantly in the middle chromosphere, shaping the moss observed above in the transition region and corona.
We identified an area with polarity reversal at the upper chromosphere around the edge of the pore, suggesting the presence of a magnetic discontinuity in the upper chromosphere. 
Transient and recurrent jet-like events are observed in this region, likely driven by magnetic reconnection.
Around the penumbral edge, we find large-scale magnetic fields corresponding to the superpenumbral fibrils seen in the upper chromosphere.
In the superpenumbral fibrils, we find Zeeman-induced linear polarization signals, suggesting the presence of a significantly inclined magnetic field, as strong as 1000~G in the upper chromosphere.

\end{abstract}

\keywords{Solar magnetic fields (1503) --- Plages (1240)  --- Spectropolarimetry (1973) --- Solar magnetic reconnection (1504)
--- Solar chromosphere (1479)}


\section{Introduction} \label{sec:intro}
Typical solar active regions consist of three main features on the photosphere: sunspots, pores, 
and plages.
Sunspots and pores appear darker than their surroundings, while plages are characterized by bright patches.
Pores are essentially small sunspots that lack a penumbra.
All of these structures are associated with magnetic fields on the order of kG in the photosphere \citep[e.g.,][]{1908ApJ....28..315H,1997ApJ...474..810M}.
In general, the magnetic field lines of sunspots and pores are nearly vertical at their centers and bend toward the outer edge in the photosphere.
The inclination of the sunspot magnetic field lines becomes more horizontal in the outer penumbra, where it becomes approximately $\sim80^{\circ}$ from the local normal \citep[][]{2011LRSP....8....4B},
whereas pores exhibit a smaller change in inclination, roughly by $\sim30^\circ$ \citep{2023A&A...674A..91C}.
On the other hand, the magnetic field lines in plages are predominantly vertical, with an average inclination of $10^{\circ}$ \citep{1997ApJ...474..810M}.
It is important to note that the magnetic field lines extend above the photosphere.

In the chromosphere, sunspots and pores are often surrounded by filamentary structures called superpenumbral fibrils which extend to a much larger distance from the sunspot than the penumbra observed in the photosphere \citep{1968SoPh....5..489L}.
They are believed to be a manifestation of plasma aligned with the magnetic field lines \citep{1971SoPh...20..298F},  although they do not seem to always trace them \citep{2011A&A...527L...8D}.
A variety of dynamic phenomena such as brightness variability, oscillatory motions, and inverse Evershed flow occur in the superpenumbral fibrils \citep{2013A&A...560A..84S,Chae_2014,2020ApJ...891..119B,2021RSPTA.37900183M} and they play a key role in the energy transfer and dissipation in the solar atmosphere \citep{2024ApJ...970...66B}.
Sunspot plumes, which can exhibit strong EUV emissions, are the most prominent features in the transition region and lower corona above sunspots \citep{1974ApJ...193L.143F}.
Sunspot plumes are often found to be associated with significant redshifts of spectral lines formed in the upper transition region, although the generation mechanism of these downflows is under debate \citep{2018GSL.....5....4T}.

Plages are more prominently visible as bright structures in the chromosphere than in the photosphere.
The thermal and dynamical properties have been investigated using strong resonance lines such as Ca~{\sc ii} H and K, and Mg~{\sc ii} $h$ and $k$ \citep[e.g.,][]{1974SoPh...39...49S}.
The intensity profile of the Mg~{\sc ii} $k$ line at $2795~\mathrm{\AA}$ observed in plages implies a hot and dense chromosphere, indicative of local chromospheric heating \citep{2015ApJ...809L..30C}.
Over the plage chromosphere in active regions, low-lying EUV emissions called moss exist and correspond to the footpoints of hot ($3-5$~MK) coronal loops \citep{1999SoPh..187..261S,1999SoPh..190..409B,1999SoPh..190..419D}.
The moss exhibits temporal variations on 10~s time scales and spatial structures of $<1$~Mm.
The location of EUV emissions in the moss does not correlate well with the locations of underlying magnetic elements in the lower chromosphere and photosphere.
Some observations suggest that the heating and dynamics of the moss is caused by magnetic field braiding, i.e., in the upper parts of hot coronal loops \citep{2005ApJ...621..498K,2013ApJ...770L...1T} as well as more local heating \citep{2024NatAs...8..697B}.

The magnetic field is responsible for the above mentioned phenomena, and thus its determination throughout the solar atmosphere is critical to understand their driving mechanism.
Studying the stratification of the magnetic field requires observing multiple spectral lines that are formed in different regions of the solar atmosphere. 
While there are spectral lines in the visible and in the infrared that form in the photosphere and in the lower and middle chromosphere, spectral lines forming in the upper chromosphere are typically found in the UV region of the spectrum, and thus out of reach of ground-based facilities. 
The CLASP sounding rocket experiments aimed at demonstrating that UV spectropolarimetry of the H~{\sc i} Lyman-$\alpha$ line \citep{2017ApJ...839L..10K} and of the Mg ~{\sc ii} $h$ and $k$ doublet \citep{2021SciA....7.8406I} are suitable for the inference of the magnetic field in the upper chromosphere. In particular, the Chromospheric LAyer Spectro-Polarimeter (CLASP2), which flew in 2019, provided the first ever spectrally and spatially resolved full Stokes spectra across the Mg~{\sc ii} $h$ and $k$ lines at three positions on the solar disk:
the disk center, an active-region plage, and a quiet region near the solar limb.
The Zeeman-induced circular polarization spectra were detected in several UV spectral lines and the longitudinal magnetic field from the lower to the upper chromosphere in the plage along a single slit was obtained \citep{2021SciA....7.8406I,2023ApJ...945..144L,2023ApJ...954..218A}.
The scattering of anisotropic radiation dominates the linear polarization spectra \citep{2012ApJ...750L..11B}, and the theoretically predicted shape of the linear polarization spectra was observationally verified \citep{2022ApJ...936...67R}.
Furthermore, the impact of the chromospheric magnetic field on the linear polarization via the Hanle and Magneto-Optical (MO) effects \citep{2016ApJ...831L..15A,2016ApJ...830L..24D} was confirmed \citep{2023ApJ...945..125I}, while the application of the HanleRT-TIC \citep{2016ApJ...830L..24D,2022ApJ...933..145L} allowed to estimate the magnetic field vector  at several positions of the slit \citep{2024ApJ...975..110L}.

In this paper, we investigate the magnetic properties of an active region from the photosphere to the upper chromosphere using data from the Hinode satellite and the CLASP2.1 sounding rocket experiment.
CLASP2.1, which is a reflight mission of the CLASP2, performed scan observations over an active region and succeeded in measuring the Stokes spectra across the Mg~{\sc ii} $h$ and $k$ lines over the different active region structures of a plage, a pore, and superpenumbral fibrils.
\cite{2024ApJ...974..154L} applied the HanleRT-TIC to the intensity and circular polarization profiles obtained by CLASP2.1 to infer the stratification of the temperature, the electron density, the line-of-sight (LOS) velocity, the microturbulent velocity, and the longitudinal component of the magnetic field.
They found that the brightness pattern of the plage region resembles the magnetic field in the chromosphere and the overlying moss, suggesting a common magnetic origin for heating in both areas.
Here, instead, we apply the weak-field approximation (WFA), which is a relatively simple but robust method to extract the magnetic field information from the polarization of spectral lines, and compare the derived magnetic properties with the plasma structures throughout the solar atmosphere from the lower chromosphere to the corona in great detail.  

\section{Observations} \label{sec:obs}
On October 8, 2021, a suborbital rocket experiment, CLASP2.1, observed the National Oceanic and Atmospheric Administration (NOAA) active region 12882 located near the solar disk center (Figure \ref{fig:sdo_hinode}a).
Coordinated observations with the Solar Optical Telescope (SOT) aboard the Hinode satellite  and the Interface Region Imaging Spectrograph (IRIS) were performed.
These coordinated observations were complemented by data acquired by the Solar Dynamics Observatory (SDO) satellite.
The details of the observations by each instrument are given in the following subsections.

The Hinode/SOT has provided high-resolution intensity images and longitudinal magnetic field maps of the underlying photosphere  (Figures~\ref{fig:sdo_hinode}{\bf c} and \ref{fig:sdo_hinode}{\bf d}), while the Atmospheric Imaging Assembly (AIA) aboard SDO provided images of the plasma structures from the low chromosphere to the corona (Figure~\ref{fig:sdo_hinode}, panels e and g$-$i).
IRIS provided high spatial resolution intensity images at the wavelengths of the Mg~{\sc ii} $h$ \& $k$ lines (i.e., the same spectral lines observed by CLASP2.1), providing information on the fine-scale structures in the chromosphere (Figure~\ref{fig:sdo_hinode}f).
CLASP2.1 observed an active region plage (the strongly magnetized region outside the sunspot). 
As shown in the photospheric intensity image (Figure~\ref{fig:sdo_hinode}c), CLASP2.1 also observed the edge of a penumbra, which is a brighter region in the sunspot and is composed of filamentary structures.
The plage is visible as bright structures in the lower chromosphere (Figure~\ref{fig:sdo_hinode}e) and exhibits strong ($>1$~kG) magnetic fields with opposite polarity to that of the sunspot (Figure~\ref{fig:sdo_hinode}d) at the photospheric layer.
In the transition region, as seen by the AIA 304~{\AA}~and 171~{\AA} channels (Figures~\ref{fig:sdo_hinode}g and \ref{fig:sdo_hinode}h), the upper part of the scanned region (roughly at $>0\arcsec$ in SJ Y) is dominated by the bright and mottled structures called the moss.
These moss regions are located at the footpoints of hot ($2-5$~MK) coronal loops, which are diffuse as observed by the AIA 94~{\AA} and 335~{\AA} channels (panels b$-$2 and i of Figure~\ref{fig:sdo_hinode}).
The lower part of CLASP2.1 observing region ($\mathrm{SJ~Y}<\sim0\arcsec$) is covered by cool ($<2$~MK) loops \citep[or sunspot plumes;][]{1974ApJ...193L.143F}, rooting from the sunspot as seen in AIA 171~{\AA} images (Figure~\ref{fig:sdo_hinode}h). These loops are also visible in AIA~304~{\AA} images (Figure~\ref{fig:sdo_hinode}g).
The $k$ core intensity image from IRIS (Figure~\ref{fig:sdo_hinode}f), which shows the plasma structure at the upper chromosphere, reveals the existence of the so-called superpenumbral fibrils, which are the filamentary structures routing closer to the sunspot center and extending longer towards the outside of the sunspot than the penumbra 
seen in the photosphere.

\subsection{CLASP2.1}
The CLASP2.1 instrument is identical to that of CLASP2 \citep{2020SPIE11444E..6WT}.
It was designed to measure the variation with wavelength $\lambda$ of the Stokes $I$, $Q$, $U$, and $V$ parameters across the Mg~{\sc ii} $h$ \& $k$ lines around 2800~\AA, where $I$ is the intensity, $Q$ and $U$ the linear polarization, and $V$ the circular polarization.
The spatial and spectral resolutions are $\sim1\farcs1$ and $\sim0.1$~\AA~\citep{2018SPIE10699E..2WS,2018SPIE10699E..30Y}, respectively.
The instrument also included the slit-jaw (SJ) monitor system, which takes fast cadence (0.6~s) images of the upper chromosphere in the Lyman-$\alpha$ passband (FWHM$=35~\mathrm{\AA}$), with a field of view (FOV) of $527\arcsec\times527\arcsec$ \citep{Kubo2016}.
The SJ images are used for the co-alignment with other observations.

\begin{figure}
	\begin{center}
		\includegraphics[angle=90, width = \textwidth]{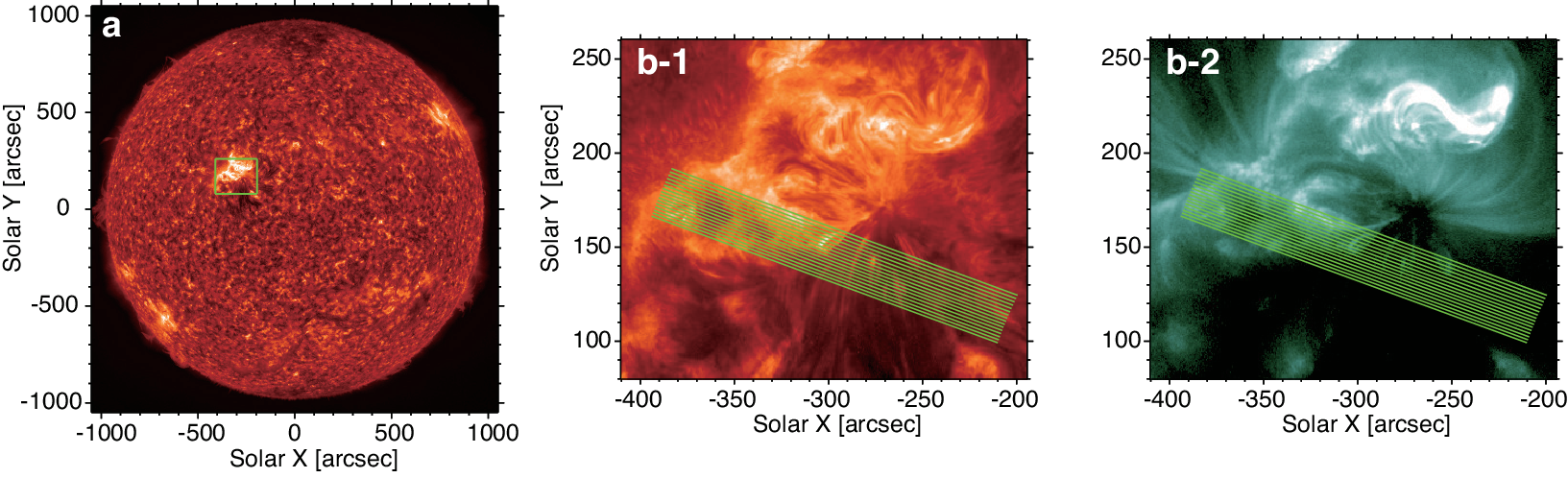}\\
		\includegraphics[angle=90,width = \textwidth]{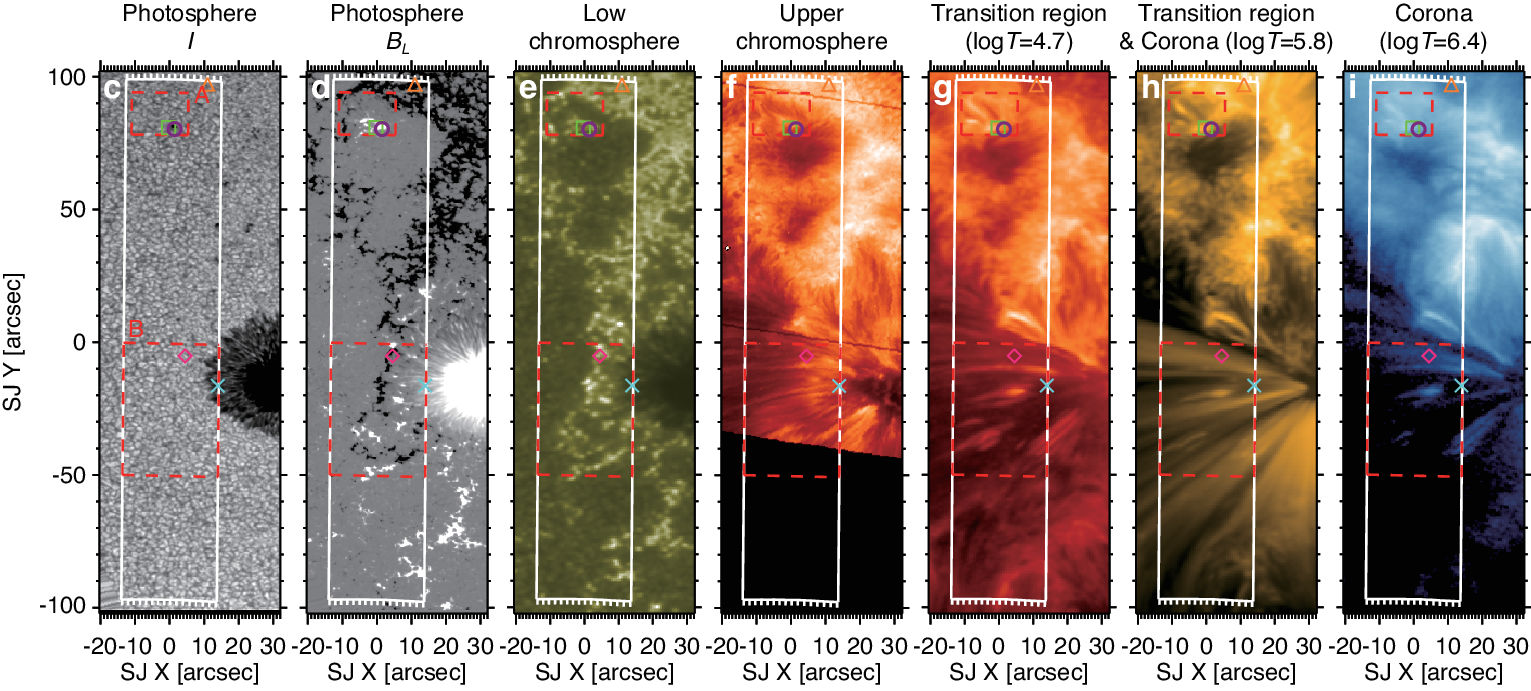}
	\end{center}
	\caption{Overview of the observed area. 
	Panels {\bf a}, {\bf b$-$1}, and {\bf b$-$2}:
	The AIA 304~{\AA} (panels a and b$-$1) and 94~{\AA} (panel b$-$2)
	images temporally averaged between 17:41.29 and 
	17:47.41 UT, and between 17:41.35 and 
	17:47.47 UT, respectively, on 2021 October 8. 
	The green box in panel a indicates the areas displayed in panels b$-$1 and b$-$2. 
	The 16 locations of the CLASP2.1
	spectrograph's slit are highlighted in green in panels b$-$1 and b$-$2. 
	The slit length is $196\arcsec$ and the scan step size is $1\farcs8$.
	Panel {\bf c}: The photospheric intensity image obtained from the continuum near the Fe~{\sc i} lines at 6302~{\AA} taken by SP of Hinode/SOT.
	Panel {\bf d}: Longitudinal component of the photospheric magnetic field ($B_{L}$) inferred from the Stokes profiles of the Fe~{\sc i} lines observed by Hinode/SOT-SP clipped to $\pm1$~kG.
	Panel {\bf e}: Intensity image of the low chromosphere observed by SDO/AIA 1600~{\AA}.
	Panel {\bf f}: Intensity image at the wavelength position of the Mg~{\sc ii}~$k$ center
	taken by IRIS/SG from 16:37:10 UT to 17:04:41 UT on 2021 October 8.
	Panel {\bf g}: SDO/AIA 304~{\AA} intensity image of the upper chromosphere and the transition region.
 	Panels {\bf h} and {\bf i}: The coronal images taken by SDO/AIA 
	at 171~{\AA} and 335~{\AA}.
	The SDO/AIA images are temporally averaged over the CLASP2.1 observing duration.
	The white box shows the region scanned by CLASP2.1 and white tick marks 
	indicate the edges of 16 locations of the spectrograph's slit.
	The dashed red rectangles indicate the regions of interest (ROIs) in our study (ROI A in Figure~\ref{fig:enlarge1}  
	and ROI B in Figure~\ref{fig:enlarge2}).
	The symbols in the c$-$i panels indicate the location of the profiles shown in
	Figure~\ref{fig:clasp21prof} (light blue cross mark) and Figure~\ref{fig:profs} (pink diamond, orange triangle, purple circle, and green square).
	}
	\label{fig:sdo_hinode}
\end{figure}

After the 16~s disk center observation acquired for calibration purposes, CLASP2.1 successfully measured the full Stokes spectra from 2793.3~{\AA} to 2806.6~{\AA} across part of the NOAA 12882 plage region at 16 adjacent locations of the $196\arcsec$ long spectrograph slit (green lines in panels b$-$1 and b$-$2 of Figure~\ref{fig:sdo_hinode}), from 17:42:13 to 17:47:38 UT.
The spatial interval between two adjacent slit locations is 1\farcs8, resulting in a scan area of 29\arcsec$\times$196\arcsec.
The total observing time for each slit position is longer than 17.6~s to obtain the Stokes $V$ spectra with a reasonably high signal-to-noise (S/N) ratio.
The continuously rotating ($3.2~\mathrm{s}/\mathrm{rotation}$) zero-order MgF$_{2}$ waveplate \citep{Ishikawa2013} of the Polarization Modulation Unit \citep[PMU;][]{2015SoPh..290.3081I}, which is located upstream of the slit, modulates the incident radiation.
The two CCD cameras record the modulated intensity of the two orthogonal polarization states every 0.2~s in synchronization with the PMU.
After the dark and gain corrections, we perform the demodulation to derive the Stokes $I$, $Q$, $U$, and $V$ spectra for each camera and for each half rotation of the PMU.
Subsequently, we correct for wavelength drifts and co-register the spectra.
Then, we apply the response matrix of the instrument, which was derived in a polarization calibration investigation using the CLASP2 data \citep{2022SoPh..297..135S}.
Finally, we obtain the $I$, $Q$, $U$, and $V$ Stokes profiles by combining all the spectra from the two cameras at each slit position.
The reference direction for positive (negative) Stokes $Q$ is parallel (perpendicular) to the nearest solar limb, while the reference 
direction for positive (negative) Stokes $U$ is at $45^{\circ}$ counterclockwise (clockwise) with respect to the $Q>0$ reference direction.
The plate scales are $0\farcs529/\mathrm{pix}$ along the slit and $49.82$~$\mathrm{m}${\AA}$/\mathrm{pix}$ in the dispersion direction. 

Examples of observed $I$ and $V/I$ spectra are shown in Figures~\ref{fig:clasp21prof} and \ref{fig:profs}.
The photon noise levels of $V/I$ per spatial pixel, wavelength pixel, and slit position around the $h$ and $k$ cores are, on average, $0.17\%$ and $0.19\%$, respectively. 
These are smaller than the observed circular polarization signals.
The $Q$ and $U$ spectra are too noisy and we only deal with them by spatially averaging, in order to get a reasonable S/N (Section~\ref{Sect:BT}).

\subsection{SDO}\label{SDO}
The AIA instrument \citep{2012SoPh..275...17L} on board the SDO satellite \citep{2012SoPh..275....3P} takes full disk images of the solar atmosphere in multiple wavelengths with a spatial plate scale of 0\farcs6.
In this work, we use the data from four EUV channels, namely, 304~\AA~(He~{\sc ii}, $\log T=4.7$), 171~\AA~(Fe~{\sc ix}, $\log T=5.8$),  335~\AA~(Fe~{\sc xvi}, $\log T=6.4$), and 94~\AA (Fe~{\sc xviii}, $\log T=6.8$), and the 1600~\AA~UV channel (C~{\sc iv} and continuum emission).
The EUV and UV channels of the AIA instrument take images with a temporal cadence of 12~s and 24~s, respectively.
The spatial resolution is $1\farcs5$.

The Helioseismic and Magnetic Imager \cite[HMI;][]{2012SoPh..275..207S} on board SDO provides full disk measurements of photospheric magnetic fields with a resolution of $1''$. We use HMI line-of-sight magnetograms recorded every 45~s.
The AIA and HMI images are calibrated and co-aligned using the data reduction procedure aia$\_$prep.pro in {\it SolarSoft} \citep{1998SoPh..182..497F}.
The AIA 304~{\AA}~channel provides images very similar to those by the CLASP2.1/SJ \citep{Kubo2016,2022A&A...657A..86G}. The SDO images are co-aligned with the CLASP2.1 data by cross-correlating the AIA 304~{\AA} and CLASP2.1/SJ images.
Figure~\ref{fig:sdo_hinode} (panels e, g$-$i) shows the SDO images co-aligned with the CLASP2.1/SJ data.

\subsection{Hinode}\label{Hinode}
The Spectro-Polarimeter \citep[SP;][]{Ichimoto2008,Lites2013} of SOT \citep{Tsuneta2008,Suematsu2008,Shimizu2008} aboard the Hinode satellite \citep{Kosugi2007}, which measures the $I$, $Q$, $U$, and $V$ profiles of the Fe~{\sc i} photospheric lines at 6301.5 and 6302.5~{\AA}, performed one raster scan from 17:01:05 to 18:05:44 UT in the fast-map mode.
In this observing mode, the spatial sampling is $0\farcs3$/pix leading to a spatial resolution of $0\farcs6$ from the Nyquist theorem, and the resulting FOV is 320\arcsec (solar east-west direction) $\times$ 150\arcsec (solar north-south direction), 
which successfully covers the area observed by CLASP2.1 (Figure~\ref{fig:sdo_hinode}c). 
The apparent longitudinal magnetic flux density, which is one of the SOT-SP Level 1.5 data products provided by the Community Spectro-polarimetric Analysis Center (CSAC) of the High Altitude Observatory (DOI: 10.5065/D6P848Z8), was cross-correlated with the temporally averaged HMI magnetogram during the CLASP2.1 observing duration for the co-alignment of the SDO and Hinode images.

We use the SOT-SP Level 2 data provided by HAO/CSAC (DOI: 10.5065/D6JH3J8D) to obtain magnetic field information in the photosphere.
The Level 2 data is the result of the inversion of the Stokes profiles using the MERLIN code based on the Milne-Eddington (ME) approximation.
Under this approximation, the magnetic field vector (field strength $B$, inclination $\theta_B$, and the azimuth $\chi_B$) is constant in the formation region of the lines.
We show the resulting LOS component of the photospheric magnetic field, $B_{L}=B\cos{\theta_B}$, in Figure~\ref{fig:sdo_hinode}d.

\subsection{IRIS}\label{IRIS}
The spectrograph (SG) of IRIS \citep{2014SoPh..289.2733D} measured the Stokes $I$ spectra in two wavelength ranges of the near ultraviolet (NUV, $2783-2835$~{\AA}) and the far ultraviolet (FUV, $1332-1407$~{\AA}).
From 17:14:30 to 18:20:41
IRIS performed continuous raster scans. 
However, the overlap with the area scanned by CLASP2.1 is too small to be useful for our analysis.
Therefore, we employ a dense raster scan acquired before the CLASP2.1 observation (from 16:37:10 to 17:04:41 UT).
The scan step size is $0\farcs35$ and the FOV covered by the SG is $112\arcsec$ 
(perpendicular to the SG's slit)$\times175\arcsec$ (parallel to the SG's slit).
The spatial and wavelength plate scales are $0\farcs33/\mathrm{pix}$ and $51~\mathrm{m}$\AA$/\mathrm{pix}$, respectively.
The IRIS and CLASP2.1 scans roughly overlap for about two-thirds of the FOV of the latter (Figure~\ref{fig:sdo_hinode}f). 
The better spatial resolution of the IRIS data makes it useful for the study of the morphological structure of the active region in the upper chromosphere.

\section{Data Analysis}
In this paper, we focus on the Stokes spectra of the Mg~{\sc ii} $h$ \& $k$ lines and the two Mn~{\sc i} resonance lines.
The former spectral lines encode information on the magnetic field in the middle and upper chromosphere, while the latter provide information in the lower chromosphere.
The $V$ spectra in these four spectral lines are induced by the Zeeman effect due to the presence of longitudinal magnetic fields.
The Stokes $Q$ and $U$ signals of the Mg~{\sc ii} $h$ \& $k$ lines are dominated by the scattering of anisotropic radiation and for strong fields there might be some signature of the Zeeman effect.
In the Mn~{\sc i} lines, the $Q$ and $U$ signals could be induced by the Zeeman effect if sufficiently strong transverse magnetic fields are present.

\subsection{Weak-Field Approximation}
In the chromosphere the magnetic field strength is generally weaker than in the photosphere whereas the temperature and the non-thermal velocity are larger.
Therefore, since the Zeeman splitting is much smaller than the width of the chromospheric spectral lines, it is often suitable to consider the weak-field approximation \citep[WFA;][]{2004ASSL..307.....L}, which is a simple and direct method to infer the magnetic field from the observed spectra.

\subsubsection{Longitudinal Magnetic Field}
In the WFA, the longitudinal component of the magnetic field, $B_{L}$, is related with the Stokes $V$ as follows: 
\begin{equation}
V(\lambda) = -\frac{e\lambda^2}{4\pi m_{e} c}g_{\mathrm{eff}}B_{L}\bigg(\frac{\partial I(\lambda)}{\partial\lambda}\bigg),
\label{eq:WFA_BL}    
\end{equation}
where 
$e$ the absolute value of the electron charge, $m_{e}$ the mass of the electron, $c$ the speed of light, $g_{\mathrm{eff}}$ the effective Land\'e factor
($4/3$ and $7/6$ for the Mg~{\sc ii} $h$ and $k$ lines, and 1.94 and 1.7 for the Mn~{\sc i} 2799.1~{\AA}~and 2801.9~{\AA}).
After calculating the wavelength derivative of the intensity, $I(\lambda)$, 
the scaling factor $B_{L}$ that provides the best fit to the $V(\lambda)$ is obtained
with the least-square minimization method.
The uncertainties in $B_{L}$ are estimated by taking into account the $\pm1\sigma$ photon noise.

\begin{figure}[ht]
	\centering
		\includegraphics[angle=90,width=\linewidth]{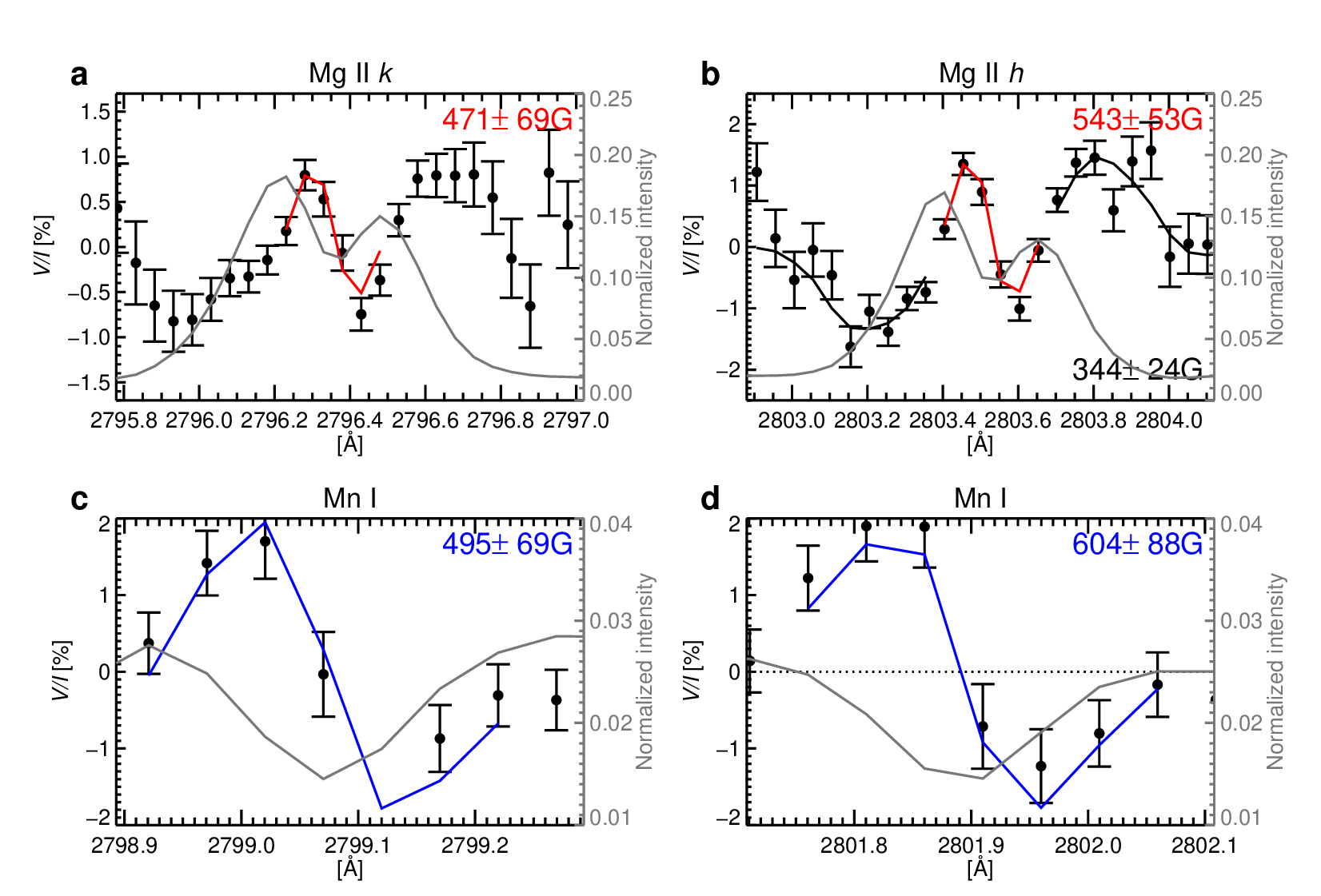}
		\caption{Normalized Stokes $I$ (gray curves) and $V/I$ (black dots) profiles of the Mg~{\sc ii} $k$ at 2796.4~{\AA} (panel {\bf a}), Mg~{\sc ii}~$h$ at 2803.5~{\AA} 
		({\bf b}), Mn~{\sc i} at 2799.1~{\AA} ({\bf c}), and Mn~{\sc i} at 2801.9~{\AA} ({\bf d}) at the location 
		in the penumbra in the photosphere, indicated by the light blue cross marks in Figure~\ref{fig:sdo_hinode} (c-i).
		The error bar shows the $\pm1\sigma$ uncertainty of $V/I$ due to the photon noise. 
		The WFA fits and the inferred $B_{L}$ values are shown in blue for the Mn~{\sc i} lines, 
		in black for the external $V/I$ lobes of the Mg~{\sc ii}~$h$, 
		and in red for the inner $V/I$ lobes of the Mg~{\sc ii}~$h$ and $k$. 
}
\label{fig:clasp21prof}
\end{figure}

We infer $B_{L}$ following the same strategy as \cite{2021SciA....7.8406I}.
Figure~\ref{fig:clasp21prof} shows the intensity and circular polarization profiles of the Mg~{\sc ii} and Mn~{\sc i} lines at the location indicated by the light blue cross mark in Figure~\ref{fig:sdo_hinode}.
As shown in the photospheric intensity image (Figures~\ref{fig:sdo_hinode}c), this pixel is located inside the penumbra, which is a brighter region of the sunspot, surrounding a dark umbra.
In most of the pixels inside the CLASP2.1 observing area, the $V/I$ profiles of the Mg~{\sc ii} $h$ and $k$ lines consist of two external lobes and two inner lobes.
The external and inner lobes of $V/I$ encode information on the magnetic field at the middle and upper layers of the chromosphere, respectively \citep{2020ApJ...891...91D}.
The WFA is valid if the longitudinal field strength is constant in the region of line formation.
It has been shown that the application of the WFA to the inner lobes of synthesized Stokes $V$
profiles from radiative MHD simulations provides a good agreement with the model value at the height where the optical depth at the line core is unity \citep{2022ApJ...936..115C,2023ApJ...954..218A}.
Therefore, we apply the WFA to the external and inner lobes separately.
The averaged values of the results from the application of the WFA to the inner lobes of the $h$ and $k$ lines are taken as the $B_{L}$ at the upper chromosphere.
Partial frequency redistribution (PRD) has an impact on the amplitudes of the external lobes \citep{2016ApJ...831L..15A,2016ApJ...830L..24D}, and the $B_{L}$ tends to be underestimated when the WFA is applied to the external lobe.
Because this effect is more significant for the $k$ line and, moreover, the blue wing of the $k$ line is blended with a Mn~{\sc i} line, we use only the external lobes of the $h$ line for the inference of $B_{L}$ at the middle chromosphere.
The $V/I$ profiles of the Mn~{\sc i} lines have two lobes that encode information at the lower chromosphere \citep{2022ApJ...940...78D} and the average value of the results from the application of the WFA to the two Mn~{\sc i} lines is taken as
the $B_{L}$ at the lower chromosphere.
The red, black, and blue lines in Figure~\ref{fig:clasp21prof} show the fits that result from the application of the WFA to the inner lobes of the $h$ and $k$ lines, to the external lobes of the $h$ line, and to the Mn~{\sc i} lines, resulting in $B_{L}=507\pm44$~G at the upper chromosphere,
$344\pm24$~G at the middle chromosphere, and $549\pm56$~G at the lower chromosphere.
Note that the WFA applied to the external lobes can underestimate $B_{L}$, and the value derived here should be considered as the lower limit.
In any case, the WFA results show that the magnetic field strength at the upper chromosphere is as high as in the lower chromosphere above the sunspot penumbra.

\begin{figure}
	\begin{center}
		\includegraphics[angle=90, width = \textwidth]{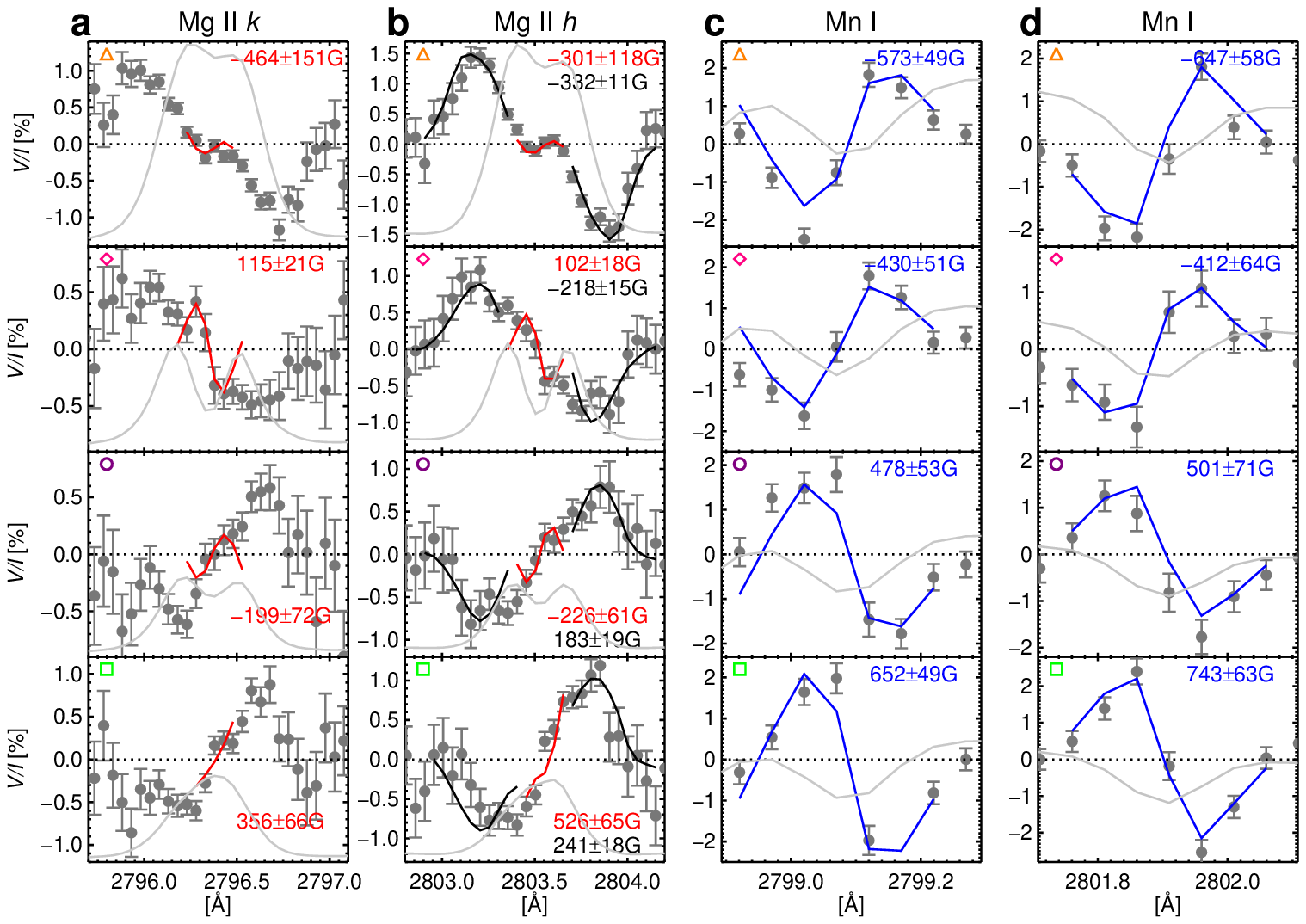}
	\end{center}
	\caption{
Examples of normalized $I$ (light gray) and $V/I$ (dark gray with $\pm1\sigma$ error bars based on the photon noise) 
		profiles at the locations marked by the orange triangle, pink diamond, purple circle, and green square in
		Figures~\ref{fig:sdo_hinode} and \ref{fig:blos} 
		for the Mg~{\sc ii} $k$ at 2796.4~{\AA} ({\bf a}), Mg~{\sc ii}~$h$ at 2803.5~{\AA} ({\bf b}),
		Mn~{\sc i} at 2799.1~{\AA} ({\bf c}), and Mn~{\sc i} at 2801.9~{\AA} ({\bf d}).
		The red, black, and blue curves show the fits resulting from the application of the WFA 
		to the inner lobes of the $h$ and $k$,
		the external lobes of the $h$, and the Mn~{\sc i} lines, respectively.
		The inferred $B_L$ values from the WFA are shown in each panel with the same color scheme.
}
	\label{fig:profs}
\end{figure}

Several examples of characteristic Stokes $V/I$ profiles and the application of the WFA 
are shown in Figure~\ref{fig:profs}.
The first row shows an example of the $V/I$ profiles observed in the unipolar plage.
The second and third rows represent the $V/I$ profiles 
showing different polarities between the upper and middle chromosphere and are discussed
in Sections~\ref{Sec:localpolchange} and \ref{Sec:superpnum}.
By applying the WFA to the inner and external lobes separately, we can successfully infer the $B_{L}$ at the upper and middle chromosphere,
which present opposite polarities, as shown by the red and black curves, respectively.
The bottom row shows an example that does not show clear $k_{3}$ or $h_{3}$ self-absorption features (see intensity profiles in the last row of Figure~\ref{fig:profs}).
In this case, the $V/I$ signals at the $4 - 5$ wavelength points around the line center are regarded as the inner lobes for the application of the WFA.
The number of pixels showing such profiles is relatively minor in the CLASP2.1 FOV.

\begin{figure}[ht]
	\centering
	\includegraphics[angle=90,width=\linewidth]{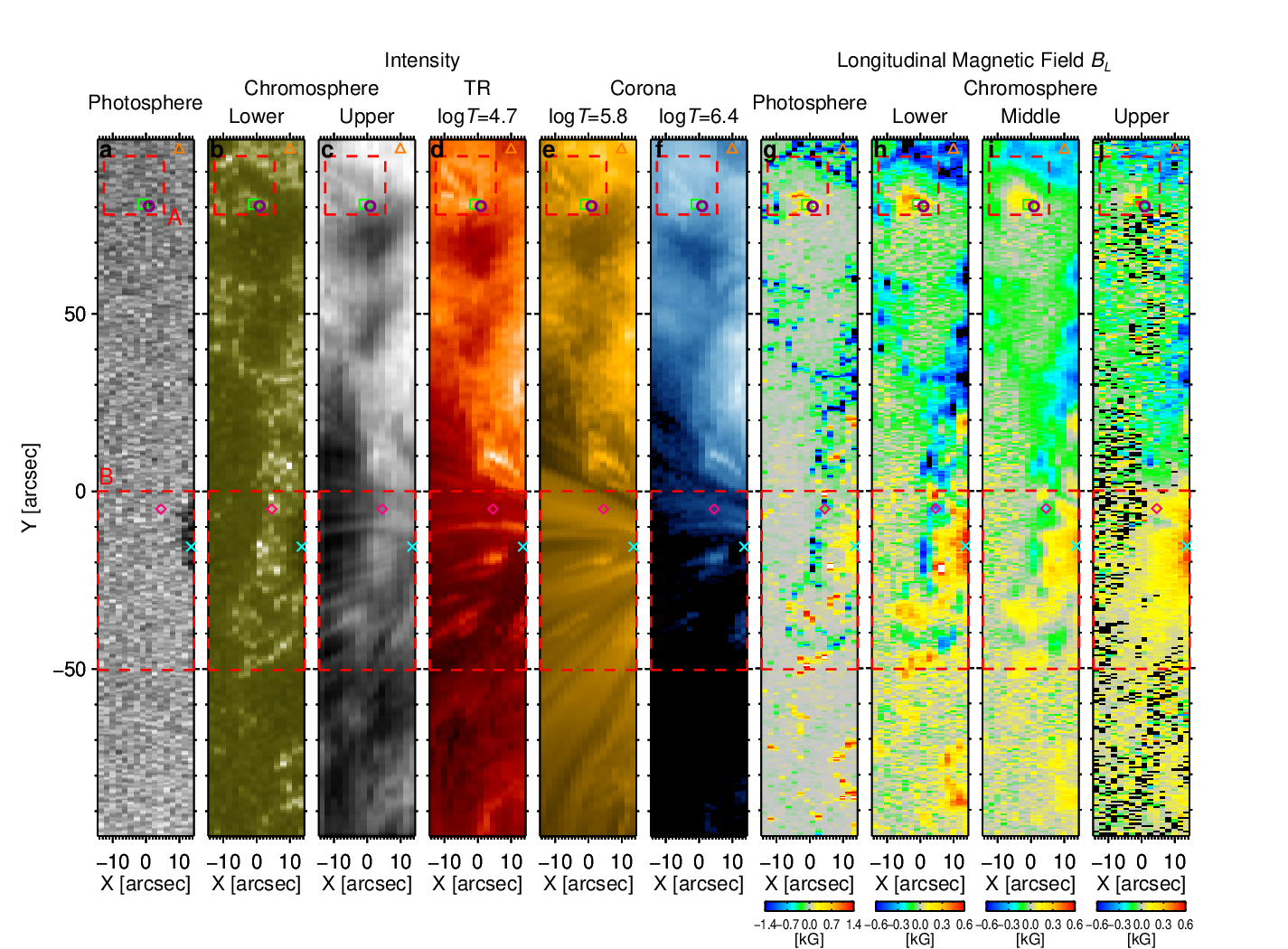}
	\caption{Panels {\bf a}$-${\bf f} and {\bf g}$-${\bf j}
	show intensity images from the photosphere to the corona, 
	and longitudinal magnetograms from the photosphere to the upper chromosphere,  
	respectively, within the CLASP2.1 scan area (indicated by white rectangles in Figure~\ref{fig:sdo_hinode}).
	{\bf a} and {\bf g} are the continuum intensity and the $B_{L}$ maps recorded by Hinode/SOT, 
	simulating the CLASP2.1 observation by picking up the signals along each CLASP2.1 slit 
	(Figs~\ref{fig:sdo_hinode}{\bf c} and \ref{fig:sdo_hinode}{\bf d}).
	{\bf b}, {\bf d}$-${\bf f} are the SDO/AIA images of 1600~{\AA} (Fig~\ref{fig:sdo_hinode}{\bf e}), 
	304~{\AA} (Fig~\ref{fig:sdo_hinode}{\bf g}), 171~{\AA} (Fig~\ref{fig:sdo_hinode}{\bf h}), 
	335~{\AA} (Fig~\ref{fig:sdo_hinode}{\bf i}) simulating the CLASP2.1 observation.
	{\bf c} is the intensity map at the core of the Mg~{\sc ii} $k$ line recorded by CLASP2.1, 
	emitted from the upper chromosphere.
	{\bf h}$-${\bf j} are the $B_{L}$ maps obtained from the WFA to the Mn~{\sc i}, 
	the external lobes of the Mg~{\sc ii} $h$, and the inner lobes of the Mg~{\sc ii} $h$ and $k$ lines 
	recorded by CLASP2.1. 
	The excluded pixels are shown in black in panel {\bf j}.
	The dashed red rectangles indicate ROIs A and B whose enlargements are shown
	in Figures~\ref{fig:enlarge1} and \ref{fig:enlarge2}.
	The symbols are common in all figures indicating the locations whose Stokes profiles are shown 
	in Figure~\ref{fig:clasp21prof} (light blue cross mark) and Figure~\ref{fig:profs} (other four marks).
	The horizontal (X) and vertical (Y) axes  are along the scan and slit directions, respectively. 
}
\label{fig:blos}
\end{figure}

Figures~\ref{fig:blos}{\bf h}$-${\bf j} show the $B_{L}$ map at the lower, middle, and upper chromosphere as obtained by applying the WFA to the Mn~{\sc i} lines, to the external lobes of the Mg~{\sc ii} $h$ line, and to the inner lobes of the Mg~{\sc ii} $h$ \& $k$ lines, respectively.
For the lower and middle chromosphere $B_{L}$ maps, the results of the WFA in all the pixels are shown.
However, for the upper chromosphere, the $V/I$ amplitudes are intrinsically smaller and the results are more affected by the noise.
We only display pixels whose maximum circular polarization amplitude across
the $h$ and $k$ lines regardless of whether it originates from inner or outer lobes,
is larger than $\pm2\sigma$ of the photon noise.

\subsubsection{Transverse Magnetic Field}
In UV spectral lines, the Zeeman-induced linear polarization is harder to detect than the 
Zeeman-induced circular polarization because it is more strongly affected by the line broadening.
Nevertheless, the CLASP2.1 observations cover a variety of magnetic structures in the observed active region, and
we aim at identifying any signature of the operation of the Zeeman effect in the linear polarization signals.
The azimuth angle of the magnetic field in the plane of sky, $\chi_{B}$ is derived by
\begin{equation}
\chi_{B} =\frac{1}{2}\arctan\bigg(\frac{U}{Q}\bigg),
\label{eq:azi}
\end{equation}
and, under the WFA,
the transverse component of the magnetic field $B_{T}$ can be related to
Stokes $Q$ as follows, when choosing a reference frame such that 
$\chi_B=0$ \citep{2004ASSL..307.....L}:
\begin{equation}
Q(\lambda_0) = -\frac{1}{4}\bigg(\frac{e\lambda^2_{0}}{4\pi m_{e} c}\bigg)^2 B_{T}^2
G_{\mathrm{eff}}\bigg(\frac{\partial^2 I}{\partial\lambda^2}\bigg), \\
\label{eq:Q_lc}
\end{equation}
\begin{equation}
Q(\lambda_w) = \frac{3}{4}\bigg(\frac{e\lambda^2_{0}}{4\pi m_{e} c}\bigg)^2 B_{T}^2
G_{\mathrm{eff}}\frac{1}{\lambda_w - \lambda_{0}}\bigg(\frac{\partial I}{\partial\lambda}\bigg), \\
\label{eq:Q_w}
\end{equation}
where $G_{\mathrm{eff}}$ is the second-order effective Land\'e factor 
($4/3$ for the Mg~{\sc ii} $h$ and $k$ lines, and 3.73 and 2.85 for the Mn~{\sc i} 2799.1~{\AA}~and 2801.9~{\AA}).
The wavelength dependence is different in the core 
($|\lambda-\lambda_0|\ll\Delta\lambda_D$, 
where $\lambda_0$ is the line-center wavelength and $\Delta\lambda_D$ the line's Doppler width) and in the line wing
($|\lambda-\lambda_0|\gg\Delta\lambda_D$).
Equations (\ref{eq:Q_lc}) and (\ref{eq:Q_w}) are used around the line center
and in the wings of the spectral line, respectively.
The relevant results are discussed in Section~\ref{Sect:BT}.

\section{Results}
\subsection{Moss Region}
At photospheric heights in the moss region,
the magnetic field is highly structured with spatial scales of a few arcseconds ($\mathrm{Y}\ge0\arcsec$, Figure~\ref{fig:blos}g).
The photospheric magnetic fields show a dominant negative polarity with significant strengths, up to 1.5~kG.
In the chromosphere, the polarities are essentially maintained but the field strength becomes much weaker than in the photosphere: $B_{L}<600$~G at the lower chromosphere and $B_{L}<400$~G at the middle and upper chromosphere (Figure~\ref{fig:blos}h$-$j).
Moreover, the magnetized region fills a much larger area in the lower/middle chromosphere than in the photosphere.
We note that the circular polarization signals in the inner $V/I$ lobes are strongly affected by noise due to their smaller amplitudes, which increases the uncertainty of the results in the upper chromosphere and makes it difficult to draw rigorous conclusions. 
In order to quantitatively evaluate how the magnetic fields expand with height,
we identify the pixels where the magnetic field strength is larger than $20\%$ of its maximum
strength in the moss region (i.e., $B_{L}>0.2\max(B_{L})$ for positive polarity  and $B_{L}<0.2\min(B_{L})$ for negative polarity) as
magnetized areas at each height 
and calculate the area ratio with respect to the one in the photosphere.
The CLASP2.1 FOV does not cover the entire moss region, and the magnetic fields could expand even outside the FOV. 
Therefore, the area ratios are derived for the whole moss region 
($\mathrm{Y}\ge0\arcsec$ but excluding the pixels where there are no significant magnetic concentrations in the photosphere at $\mathrm{X}\le-5\farcs4$, from 1st to 6th scans, and $0\arcsec\le\mathrm{Y}\le31\farcs5$)
as well as the small region which harbors  the two isolated magnetic concentrations in the photosphere at
$(\mathrm{X},\mathrm{Y})\approx(-5\arcsec, 35\arcsec)$ and $\approx(-7\arcsec,43\arcsec)$.
The details of the identified magnetized areas are shown in Appendix~\ref{AD_D} (Figure~\ref{fig:bl_area}).
Both regions show similar trends:
the magnetized region in the moss expands rapidly by a factor of 2.4 (small region) and 2.7 (whole moss region) at the 
lower chromosphere, and by a factor of 3.1 (both regions) at the middle chromosphere, compared to the photosphere.

The spatially expanded magnetized area in the middle chromosphere appears to trace the moss regions that are clearly visible in the AIA 171~{\AA} channel (Figure~\ref{fig:blos}e). 
The spatial correspondence can be confirmed by the correlation coefficients between the $B_{L}$ and the intensity (Table~\ref{tab:cc}).
The intensity of the AIA 171~{\AA} channel shows the highest correlation coefficient of 0.74 with the $B_{L}$ at the middle chromosphere, among all the combinations of wavelength channels and $B_{L}$.
The core intensity of the $k$ line and the AIA 304~{\AA} intensity also show the highest correlation coefficients with the  $B_{L}$ at the middle chromosphere, while the intensity of the AIA image at 1600~{\AA} shows the highest correlation coefficient with the $B_{L}$ at the lower chromosphere.
It is also interesting to note that the $B_{L}$ at the middle chromosphere shows a relatively high correlation coefficient with the intensity of the AIA 335~{\AA} channel, which shows the high temperature coronal plasma ($\log T=6.4$).
Again, the derived $B_{L}$ at the upper chromosphere is too noisy and the correlation coefficients are not reliable enough to be discussed. 

\begin{table}[ht]
\centering
\begin{tabular}{c|ccccc}
 & \multicolumn{5}{c}{Correlation coefficient with intensity}\\
$B_{L}$ & 1600~{\AA} & $k$ core & 304~{\AA} & 171~{\AA} & 335~{\AA} \\
\hline
Upper chromosphere & 0.25 & 0.39 & 0.31 & 0.46 & 0.34\\
\hline
Middle chromosphere & 0.66 & {\bf 0.72} & {\bf 0.65} & {\bf 0.74} & {\bf 0.63}\\
\hline
Lower chromosphere & {\bf 0.73} & 0.66 & 0.60 & 0.67 & 0.56\\
\hline
Photosphere & 0.66 & 0.46 & 0.43 & 0.46 & 0.38 \\
\hline
\end{tabular}
\caption{\label{tab:cc} Correlation coefficients between the longitudinal component of the magnetic field, $B_{L}$, at four different heights in the solar atmosphere, and the intensity from the AIA 1600~{\AA}, CLASP2.1 $k$ core, AIA 304~{\AA}, 171~{\AA}, and 335~{\AA}. The linear Pearson coefficients are calculated considering only the pixels at $\mathrm{Y}\ge0\arcsec$ in Figure~\ref{fig:blos}.
The largest correlation coefficients for each intensity are shown in boldface.}
\end{table}

\subsection{Local Polarity Reversal}\label{Sec:localpolchange}
\begin{figure}
\begin{center}
	\includegraphics[angle=90, width=\textwidth]{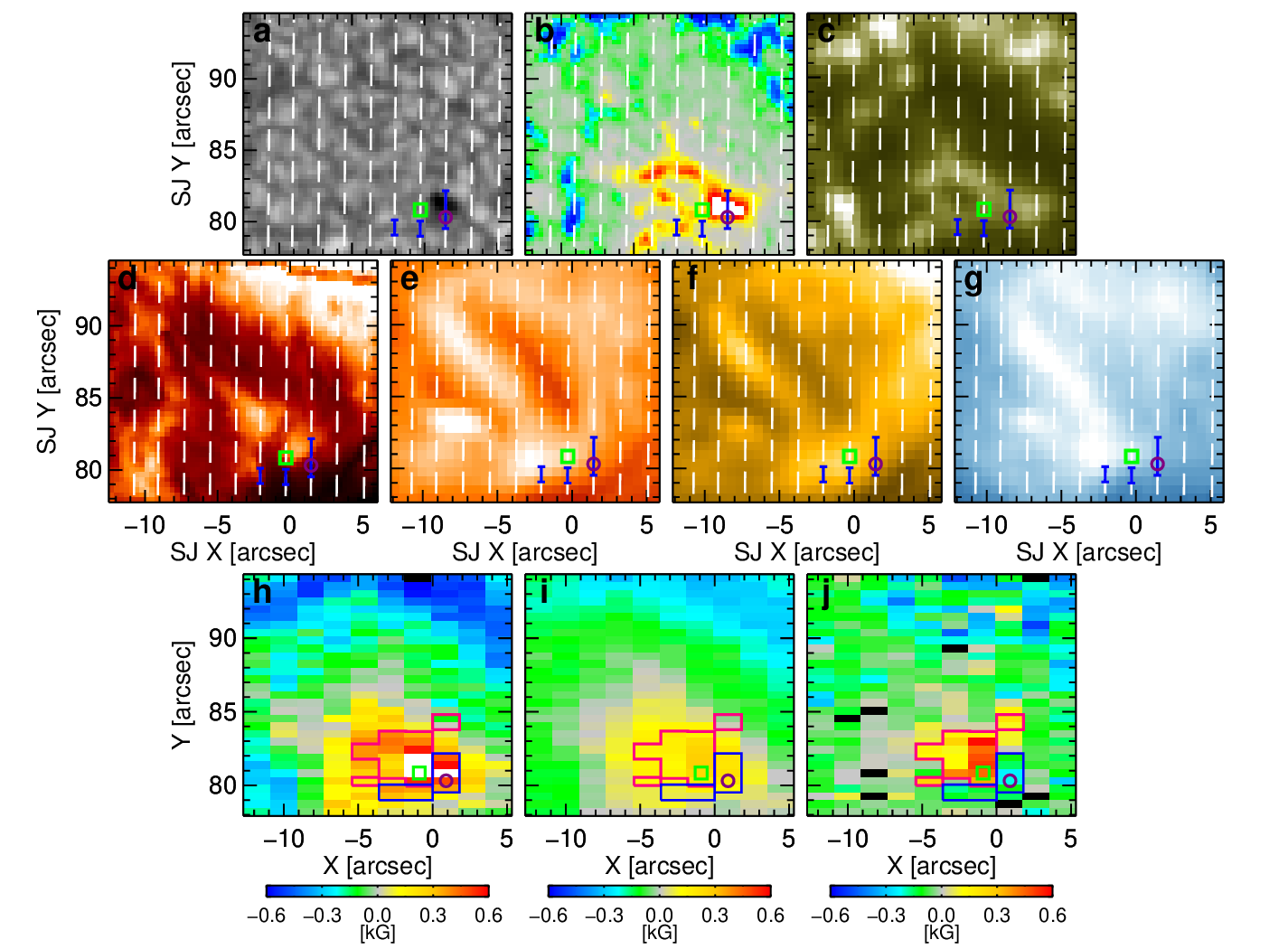}
\end{center}
\caption{ROI A.
{\bf a}: The photospheric intensity image taken by Hinode/SOT.
{\bf b}: Longitudinal component of the photospheric magnetic field observed by Hinode/SOT. The colorbar is capped at $\pm1.5$~kG.
{\bf c}: Intensity image in SDO/AIA channels of 1600~{\AA}.
{\bf d}: Intensity image at the line center of the Mg~{\sc ii} $k$ line by IRIS/SG.
{\bf e}$-${\bf g}: Intensity image in SDO/AIA channels of 304~{\AA}, 171~{\AA}, and 335~{\AA}, respectively.
The horizontal and vertical axes, SJ~X and SJ~Y, in panels {\bf a}$-${\bf g} are in CLASP2.1/SJ coordinate system as in Figure~\ref{fig:sdo_hinode}.
The blue bars indicate the locations enclosed by the blue contours in panels {\bf h}$-${\bf j}.
The white dashed lines in panels {\bf a}$-${\bf g} indicate the slit positions displayed in panels~{\bf h}$-${\bf j}.  
{\bf h}, {\bf i}, and {\bf j}: $B_{L}$ maps of the ROI A 
(red dashed boxes in Figure~\ref{fig:blos}) at the lower, middle and upper chromosphere. 
The red and blue contours are common in the bottom three panels.
The red contour indicates the area where $B_{L}>100$~G with the positive polarity at the upper chromosphere,
while the blue contour indicates the negative $B_{L}$ area at the upper chromosphere at the periphery 
of the positive polarity area (the pixels with relatively strong $B_{L}$ are visually chosen).  
The $I$ and $V/I$ profiles at the pixels indicated by the purple circle and the green square are shown in Figure~\ref{fig:profs}.
\label{fig:enlarge1}}
\end{figure}

\begin{figure}[ht]
\centering
\includegraphics[width=\linewidth]{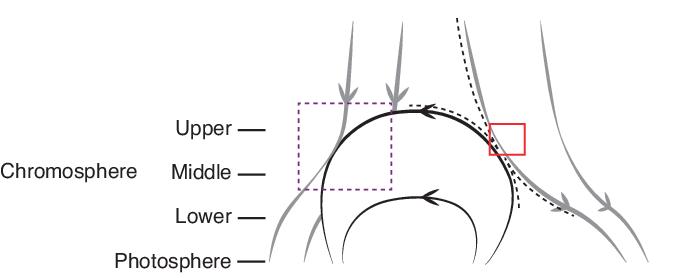}
\caption{Schematic of ROI A.
}
\label{fig:cartoon}
\end{figure}

In the moss region, the polarity is generally maintained from the photosphere to the upper chromosphere.
However, we find a location with a polarity reversal at the upper chromosphere.
Figure~\ref{fig:enlarge1} shows a zoom of the upper box of Figure~\ref{fig:blos} (ROI A).
As can be seen from the photospheric intensity image (Figure~\ref{fig:enlarge1}a),
the positive polarity region harbors a pore (dark region without penumbra).
The positive polarity region spatially expands in the lower and middle chromosphere (Figures~\ref{fig:enlarge1}h and \ref{fig:enlarge1}i).
However, at the upper chromosphere, the area of the positive polarity region is smaller and the edge shows negative polarity, as indicated by the blue boxes in Figure~\ref{fig:enlarge1}j. 
The magnetic field strength is significant, $\sim200$~G at the upper chromosphere, supporting the conclusion that the obtained negative field does not result from the failure of the application of the WFA
\citep[see also the results from the application of the HanleRT-TIC in][]{2024ApJ...974..154L}.
The third row of Figure~\ref{fig:profs} shows the $I$ and $V/I$ profiles for one of the pixels (indicated by the purple circle) that exhibits $B_{L}$ with negative polarity only at the upper chromosphere.
The WFA successfully fits the $V/I$ profiles, giving $B_{L}=-212\pm47$~G at the upper chromosphere (average of $h$ and $k$ cores) and $B_{L}=183\pm19$~G at the middle chromosphere.
In the lower chromosphere, the magnetic field strength is significantly stronger with $B_{L}=490\pm44$~G (the average of the results from the two Mn~{\sc i} lines).
A change in the area of the positive polarity region with height 
(i.e., the increase in magnetized area from the photosphere to the middle chromosphere 
and the decrease in magnetized area in the upper chromosphere)
indicates that
the magnetic field lines originating from the positive polarity region expand 
in the lower and middle chromosphere, and reach above the upper chromosphere connecting to the surrounding magnetic field patches with negative polarity (field lines in black in Figure~\ref{fig:cartoon}).
The polarity reversal at the upper chromosphere can be interpreted as  the presence of overlaying expanding plage magnetic fields with negative polarity  (red box in Figure~\ref{fig:cartoon}).

To examine whether this proposed magnetic field structure is plausible, we compare the $B_{L}$ maps with plasma structures observed by the other instruments.
The locations corresponding to the pixels showing the polarity reversal at the upper chromosphere
(blue boxes in Figure~\ref{fig:enlarge1}h$-$j) are indicated
by the blue bars in the Hinode/SOT, IRIS/SG, and SDO/AIA images (Figures~\ref{fig:enlarge1}a$-$g).
In general, the transition region and corona observed in the AIA 304~{\AA}, 171~{\AA}, and 335~{\AA} channels (Figures~\ref{fig:enlarge1}e$-$g) show similar spatial distributions of bright structures in the ROI. 
However, in contrast to the AIA 304~{\AA} and 171~{\AA} images, the bright structures do not exist in the AIA  335~{\AA} image at the location where the polarity reversal at the upper chromosphere is observed (blue bars in the figure)
and at its vicinity ($0\arcsec<$SJ~X$<5\arcsec$ and $80\arcsec<$SJ~Y$<90\arcsec$). 
The enhanced emission in the EUV can be associated with small-scale loops connecting
magnetic flux concentrations of nearby opposite polarities \citep[e.g.,][]{2019LRSP...16....2M}.
The absence of hot coronal plasma in the AIA  335~{\AA} 
channel suggests that the magnetic field lines rooted in the photospheric magnetic field concentration with positive
polarity do not extend in this region.
Instead, such field lines  
are connected along the bright structures seen in AIA 335~{\AA},
the opposite end of which appear to be rooted in the photospheric magnetic concentrations
with negative polarity, i.e., from $(\mathrm{SJ~X},\mathrm{SJ~Y})\approx(0\arcsec, 80\arcsec)$ to $(\mathrm{SJ~X},\mathrm{SJ~Y})\approx(-10\arcsec,90\arcsec)$.
The $k$ core intensity image obtained by IRIS/SG also detects the bright 
structure elongated from $(\mathrm{SJ~X},\mathrm{SJ~Y})\approx(-2\arcsec, 83\arcsec)$ to  $(\mathrm{SJ~X},\mathrm{SJ~Y})\approx(-8\arcsec,90\arcsec)$, although the IRIS data was taken roughly one hour before the CLASP2.1 observation and the detailed structure could be different.

\begin{figure}[ht]
\centering
\includegraphics[width=0.9\linewidth]{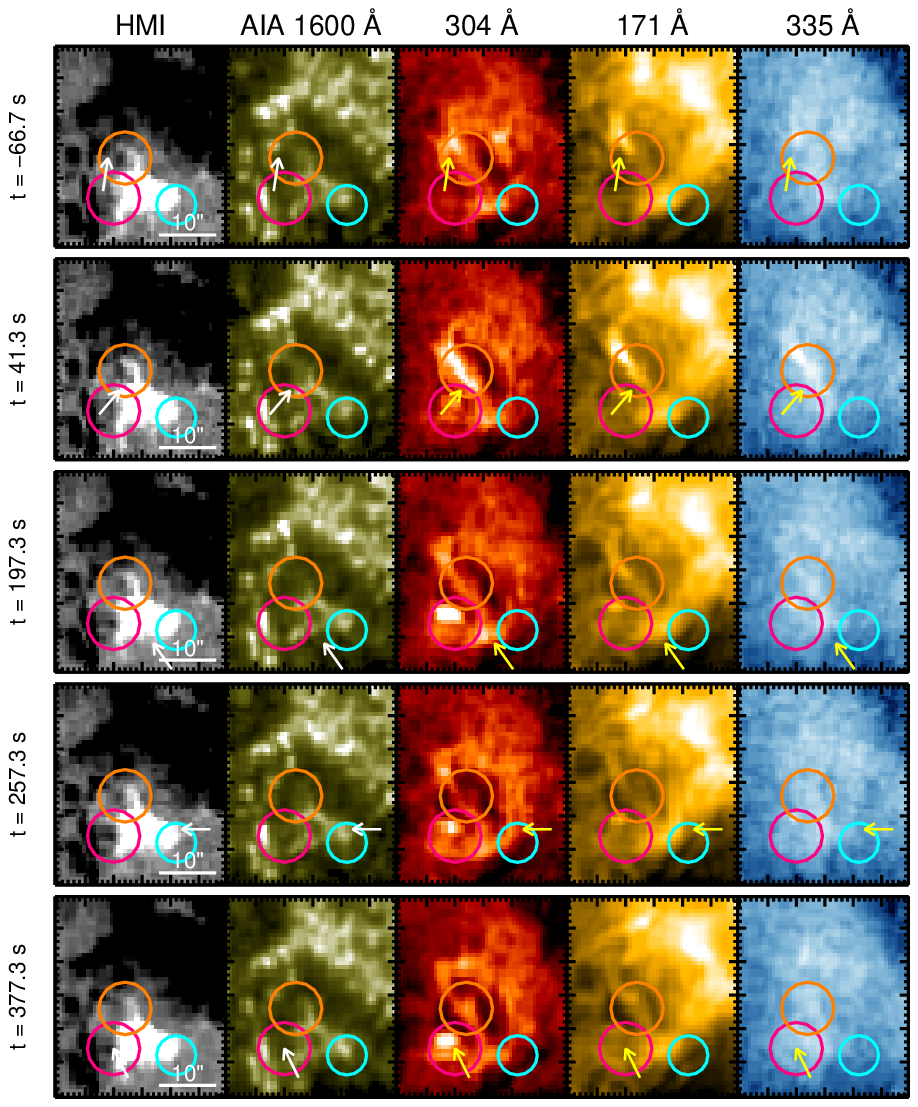}
\caption{
Temporal evolution of ROI A.
From left to right, photospheric longitudinal magnetic field $B_{L}$ obtained by SDO/HMI, 
and the intensity in 1600~{\AA}, 304~{\AA}, 171~{\AA}, and 335~{\AA} by SDO/AIA
around ROI A.
The time denoted on the left at each panel shows the time with respect to the start of the CLASP2.1 scan  
at 17:42:13 UT.
The middle three rows are during the CLASP2.1 scans ($\mathrm{t}=0-325$~s).
The light blue circles show the approximate location, specifically, the edge of the positive polarity patches, where a polarity reversal is observed in CLASP2.1.
The orange and pink circles show the locations where the cancellation 
(the negative magnetic patch, situated near the large concentration of positive polarity, gradually weakens and ultimately vanishes in the final frame)
and emergence (the magnetic patch exhibiting negative polarity becomes clearly visible in the third and subsequent frames)
of the photospheric magnetic fields are found.
The arrows show the locations of examples of the transient brightening in EUV.
A supplementary movie is also available.
}
\label{fig:sdo_rga}
\end{figure}

The presence of small-scale ($\sim$$10\arcsec$) bipolar loops indicates discontinuities of the magnetic fields in the chromosphere, which can cause reconnection with the overlying expanding magnetic fields \citep[Figure~\ref{fig:cartoon};][]{2012ApJ...751..152J}.
Even above the footpoint with the negative polarity (the purple, dashed-line box in Figure~\ref{fig:cartoon}), 
the reconnection could happen because the field lines are not perfectly parallel \citep[3D configuration of component reconnection; e.g.,][]{2012ApJ...761...87N}.
In fact, ROI A is highly dynamic in the transition region and the corona.
Figure~\ref{fig:sdo_rga} shows the temporal evolution of ROI A over more than 7 min including the CLASP2.1 observation.
The light blue circle indicates the approximate location where the polarity reversal at the upper chromosphere is detected.
The intensity originating from the chromosphere in SDO/AIA 1600~{\AA}
does not show a significant temporal variation, while
the transition region and coronal intensities (SDO/AIA 304~{\AA} and 171~{\AA} channels) do show recurrent brightening and plasma eruptions (yellow arrows) over the small-scale loops connecting the positive and negative fields.
Some of these phenomena are also identified in 
the high-temperature corona (SDO/AIA 335~{\AA}).
These dynamic activities would be partially caused by the emergence (pink circle in Figure~\ref{fig:sdo_rga})
and cancellation  (orange circle in Figure~\ref{fig:sdo_rga}) of magnetic elements
observed by the HMI photospheric magnetogram.
In the rest of the areas, there is no noticeable change in the 
photospheric magnetic field.
The brightening events at $\mathrm{t}=197.3$~s and $\mathrm{t}=257.3$~s occur at a location away from the region where the significant photospheric magnetic field changes are identified, and instead closer to the region where the polarity reversal is observed in the upper chromosphere.

\subsection{Superpenumbral fibril region}
\label{Sec:superpnum}
\begin{figure}[ht]
\centering
\includegraphics[width=\linewidth]{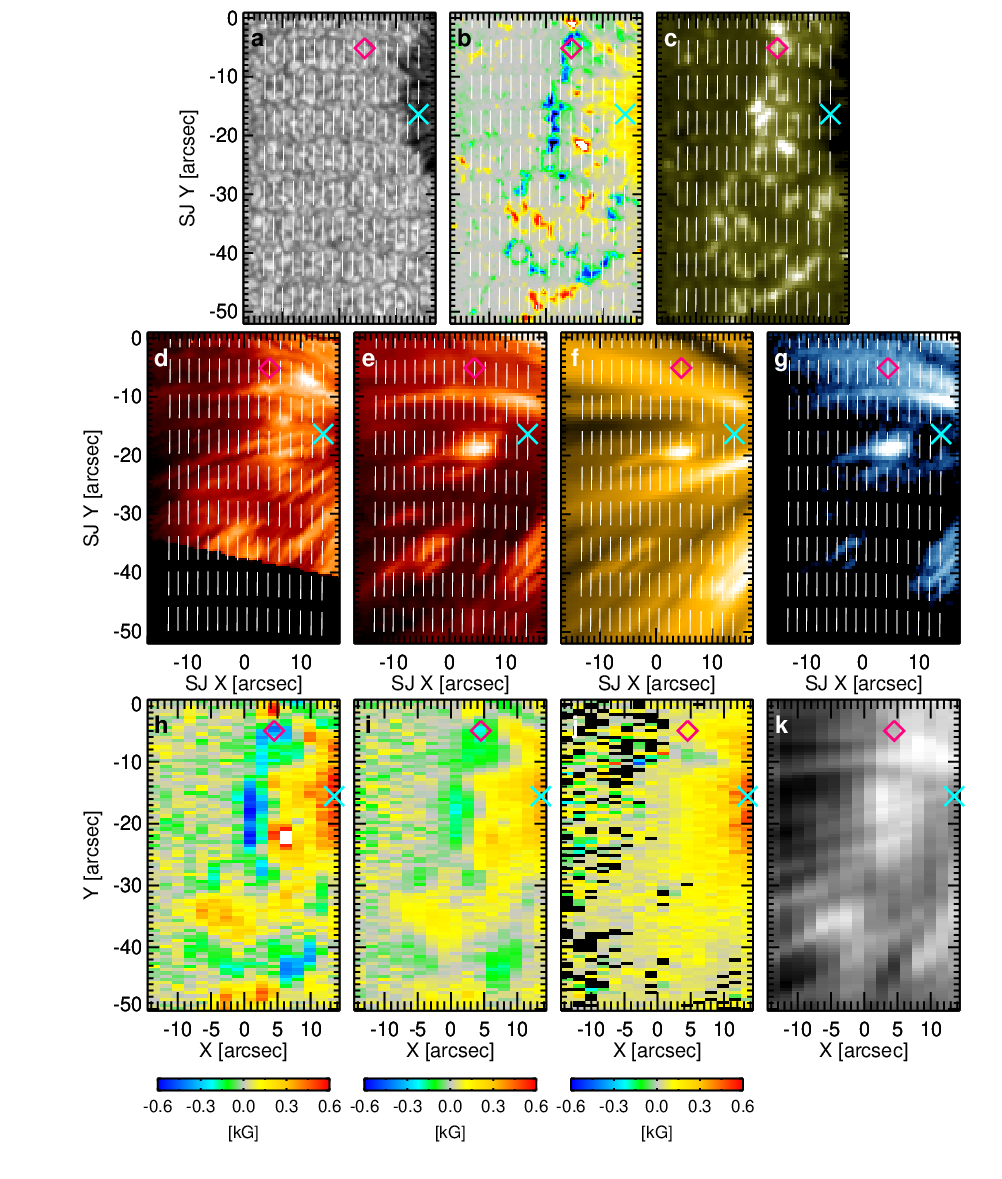}
\caption{ROI B.
{\bf a}: The photospheric intensity image taken by Hinode/SOT.
{\bf b}: Longitudinal component of the photospheric magnetic field observed by Hinode/SOT. The colorbar is capped at $\pm1.5$~kG.
{\bf c}, {\bf e}$-${\bf g}: Intensity images in SDO/AIA channels of 1600~{\AA}, 304~{\AA}, 171~{\AA}, and 335~{\AA}, respectively.
{\bf h}, {\bf i}, and {\bf j}: $B_{L}$ maps of the ROI B
(lower red box in Figure~\ref{fig:blos})
at the lower, middle, and upper chromosphere, respectively. 
{\bf d} and {\bf k}: Intensity images from the core of the Mg~{\sc ii} $k$ line 
from the IRIS/SG (Figure~\ref{fig:sdo_hinode}f) and from the CLASP2.1 (Figure~\ref{fig:blos}c), respectively.
The white dashed lines in panels {\bf a}$-${\bf g} indicate the slit positions displayed in panels~{\bf h}$-${\bf k}.  
}
\label{fig:enlarge2}
\end{figure}

\begin{figure}[ht]
\centering
\includegraphics[width=\linewidth]{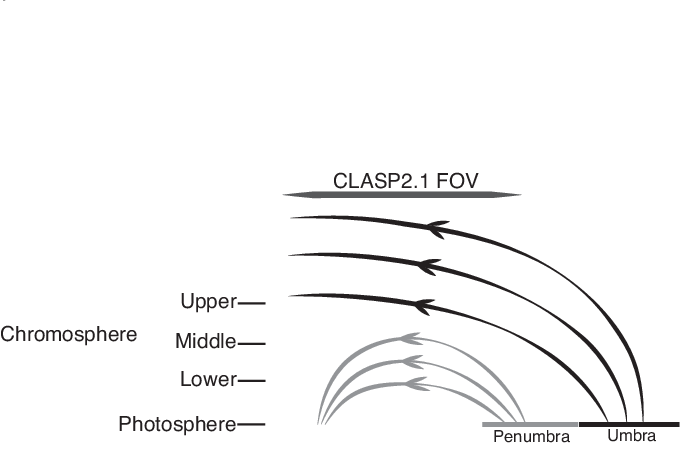}
\caption{Schematic of ROI B.
}
\label{fig:cartoon_sunspot}
\end{figure}

The CLASP2.1 FOV includes both, the edge and the outer area of the penumbra  (ROI B, lower box in Figure~\ref{fig:blos}).
At the outer periphery of the penumbra, there are magnetic patches with negative polarity, which is opposite to that of the sunspot, as observed in the photospheric $B_{L}$ map (Figure~\ref{fig:enlarge2}{\bf b}). 
These negative polarity patches are also present in the lower and middle chromosphere. 
The longitudinal field strength in the lower and middle chromosphere is weaker than in the photosphere, and the area covered by these negative polarity patches is larger (compare Figures~\ref{fig:blos}g and \ref{fig:blos}h), 
resulting in the absence of a gap between negative and positive polarity regions in the chromosphere.
In the upper chromosphere above region of the lower atmosphere characterized by negative polarity magnetic fields, no corresponding negative-polarity structures are detected; instead, the magnetic topology is dominated by magnetic fields of positive polarity.
An exception is the region around $\mathrm{X}=5\arcsec$ and $\mathrm{Y}=-9\arcsec$,
where a negative polarity field is also detected in the upper chromosphere.
The dominance of positive fields at the upper chromosphere indicates that  the majority of the large-scale magnetic fields originating from the sunspot (Figure~\ref{fig:cartoon_sunspot})
do not return to the lower atmosphere within the CLASP2.1 FOV,  at least approximately $\sim50\arcsec$ from the sunspot center.
The longitudinal field strength at the upper chromosphere is $\sim100$~G or less.  

The footprint in the intensity of these large-scale magnetic fields at the upper chromosphere would be the superpenumbral fibrils seen in the core of the $k$ line (Figures~\ref{fig:enlarge2}d and \ref{fig:enlarge2}k).
The superpenumbral fibrils extend across the CLASP2.1 FOV, suggesting the presence of large-scale magnetic fields at the upper chromosphere.
Several loops similar to these fibrils are found in the transition region and the low-temperature 
corona (Figure~\ref{fig:enlarge2}e and \ref{fig:enlarge2}f).
This similarity supports the idea that part of the magnetic fields observed in the upper chromosphere reach the transition region and the corona without returning to the lower atmosphere.
Contrary to the moss region, no significant correlation is found between the intensity and the $B_{L}$ (Appendix~\ref{ap:cor}).
The exception is the correlation between the intensity in the AIA 1600~{\AA} channel and the $B_{L}$ at the photosphere: the moss region and the fibril region show similar correlation coefficients of $\sim0.6$.
The filamentary structure of the fibrils seen in the $k$ line core intensity is not noticeable in the $B_{L}$ map at the upper chromosphere.
The relatively strong $B_{L}$ region does not extend over the CLASP2.1 FOV and highly inclined magnetic fields would dominate the superpenumbral fibrils.

\subsection{Upper Limit of $B_{T}$}\label{Sect:BT}

\begin{figure}[ht]
\centering
\includegraphics[angle=90,width=\linewidth]{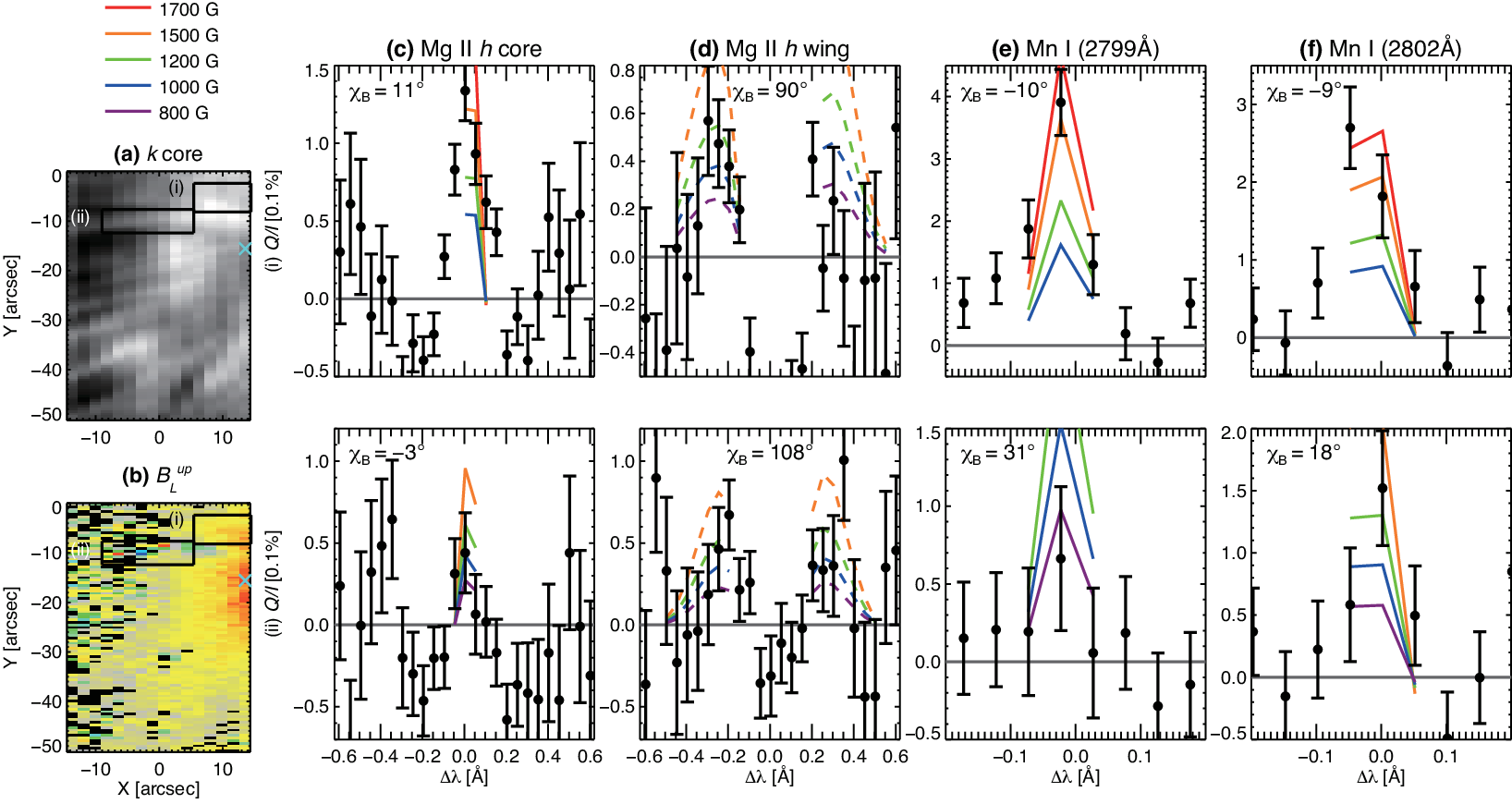}
\caption{Panels {\bf (a)} and {\bf (b)} are the images of the $k$ core intensity and $B_{L}$ at the upper chromosphere of ROI B, respectively.
The black boxes show the areas (i) and (ii) over which the Stokes profiles are averaged.
Panels {\bf (c)}$-${\bf (f)} are the spatially averaged $Q/I$ profiles in the unit of 0.1~\% around 
the Mg~{\sc ii} $h$, the Mn~{\sc i} at 2799~{\AA}, and the Mn~{\sc i} at 2802~{\AA}.
The reference direction of linear polarization is chosen such that $\chi_B$=0 at the line center for (c), (e), and (f), and $\chi_B$=0 at the blue wing ($-0.3~\mathrm{\AA}\le\Delta\lambda\le-0.2~\mathrm{\AA}$) of the $h$ line for (d).
The azimuth angles $\chi_B$, which are derived from the original $Q$ and $U$ signals using Equation~(\ref{eq:azi}) and are used for changing the reference frame, are shown in each panel.
The error bars show the $\pm1\sigma$ photon noise.
The upper row is for area (i), while the lower row is for area (ii).
The color lines show the $Q/I$ amplitudes calculated with the weak-field approximation for the different transverse magnetic field $B_{T}$ (red, orange, green, blue, and purple for $B_{T}=1700,$ $1500,$ $1200,$ $1000,$ and $800$~G, respectively).
The solid and dashed line are the estimated $Q/I$ signals for the line center and the wings, respectively.
}
\label{fig:av_prof_rg2}
\end{figure}

Here, we try to get some hints about the presence of transverse fields from the linear polarization profiles $Q$ and $U$. 
The $Q$ and $U$ profiles from a single pixel are too noisy, and thus we spatially average the profiles over the areas where the magnetic structure is expected to be similar (black boxes in Figure~\ref{fig:av_prof_rg2}):
area (i) around the penumbral region where relatively strong transverse magnetic fields exist, at least in the photosphere and lower chromosphere, and area (ii) corresponding to conspicuous superpenumbral fibrils observed in the $k$ core intensity.

Theoretical calculations in plane-parallel models of the solar atmosphere show that the linear polarization signals at the $k$ line are dominated by the scattering of anisotropic radiation \citep{2012ApJ...750L..11B}.
For a close-to-limb geometry, the $Q/I$ amplitudes are of the order of a few \% at the near wings and $\sim$1\% around the line center. 
Similar $U/I$ patterns can be induced by the operation of the Hanle effect in the core and by the Magneto-Optical effects in the wings \citep{2016ApJ...831L..15A,2016ApJ...830L..24D}.
Our observing target is close to the disk center and the amplitudes are not as significant as near the limb. 
However, the lack of axial symmetry of the incident radiation field 
can significantly contribute to the linear polarization signals \citep[and references therein]{2023ApJ...945..125I}.
Indeed, in both areas, the observed linear polarization signals around the $k$ line exhibit signatures of scattering polarization, and therefore the Zeeman-induced linear polarization signals of the $k$ line itself are not investigated in this study.
However, we investigate the impact of the Zeeman effect on the Mg~{\sc ii} $h$ and the Mn~{\sc i} lines, which are insensitive to the Hanle effect.

Figure~\ref{fig:av_prof_rg2} shows the spatially averaged linear polarization profiles of the Mg~{\sc ii} $h$ line and the two Mn~{\sc i} lines in areas (i) and (ii).
The azimuth angles ($\chi_B$), derived from the line center $Q$ and $U$ signals using Equation (\ref{eq:azi}), are indicated in panels c, e and f.
We also obtain $\chi_B$ from $Q$ and $U$ signals at the blue wing ($-0.3~\mathrm{\AA}\le\Delta\lambda\le-0.2~\mathrm{\AA}$) of the $h$ line (panel d).
These angles $\chi_B$ are used to rotate the reference frame of the Stokes parameters such that the $U$ signal at the line center of each spectral line becomes zero 
and the $Q$ signal becomes positive (i.e., $\chi_B=0$ in the new reference frame).
Therefore, we do not show the $U/I$ profiles here.
$\chi_B=0^{\circ}$ corresponds to the direction perpendicular to the solar radial direction (approximately perpendicular to the slit direction).
The $\chi_B$ is different between the core and the blue wing of $h$ line and thus the resulting $Q/I$ profiles look different as shown in panels c and d.
In panels c, e, and f, the $Q/I$ signals at the central three pixels of the line center, calculated using 
Equation (\ref{eq:Q_lc}) for different transverse magnetic field strengths,
are shown in color. 
For the wing of the $h$ line (panel d), we show
the $Q/I$ signals at the seven pixels for each wing ($|\Delta\lambda|\ge0.2$~\AA)
based on  Equation  (\ref{eq:Q_w}).

Concerning the impact of the scattering of anisotropic radiation on the $h$ line,
the signal at the line center is theoretically predicted to be zero\footnote{The Mg~{\sc ii} $h$ line is a transition between levels with $J=1/2$. Therefore, there is no scattering polarization at its line center.} with antisymmetric features of the order of 0.1~\% around the line center \citep{2012ApJ...750L..11B}. 
In both areas, non-zero $Q/I$ signals, which exceeds the noise level,
are detected at the  core of the $h$ line,
and a clear antisymmetric shape is not identified.
Assuming that the detected $Q/I$ signals at the line center originate from the Zeeman effect, the transverse magnetic field strength in the upper chromosphere would exceed 1500~G with $\chi_B=11^{\circ}$ and 1000~G with $\chi_B=-3^{\circ}$ in areas (i) and (ii), respectively, as the $Q/I$ amplitudes at the line center in areas (i) and (ii) are comparable to or larger than the estimated Zeeman-induced signals with $B_{T}=1500$~G (orange line) and $B_{T}=1000$~G (blue line), respectively.
On the other hand, it is difficult to assess the presence of the Zeeman effect based on the polarization signals in the wings of the $h$ line.
As shown in panel d, both areas show a clear positive $Q/I$ signals in blue wings,
while they do not in the red wings. 
The discrepancy between the red and blue wings may indicate the limitation of the weak-field approximation, namely that magnetic field is not constant within the formation region of the wing. Additionally, the near-wing $Q/I$ signals of the $h$ line could be influenced by scattering polarization.

In area (i), the presence of Zeeman signals in the $h$ core is supported by 
the Zeeman-induced signals observed in the Mn~{\sc i} lines.
The $Q/I$ signals in both Mn~{\sc i} lines exhibit amplitudes that suggest  $B_{T}$ values between $1500$~G (orange lines in panels d and e) and $1700$~G (red lines in panels d and e) with consistent azimuth angles of $\chi_B\sim-10^{\circ}$.
The longitudinal field strength ($B_{L}$) in the upper chromosphere is approximately 170~G, 
indicating that the magnetic field is nearly horizontal in the upper chromosphere.
In area (ii), however, the $Q/I$ signals in the Mn~{\sc i} lines are inconsistent with each other, showing different line-core amplitudes and azimuth angles $\chi_B$, indicating the absence of clear transverse magnetic fields in the lower chromosphere after
spatially averaging.
The difference in the $Q/I$ signals between the $h$ core and the two Mn~{\sc i} lines is consistent with
the idea that the magnetic structures observed in the $h$ core, representing large-scale magnetic field lines, only exists in the upper chromosphere (Figure~\ref{fig:cartoon_sunspot}).
Moreover, the value of $\chi_B=-3^\circ$ derived from the $h$ line core is roughly aligned with the the observed fibrils along the $\mathrm{X}$ axis, suggesting that the $h$ line core signals originate from the Zeeman effect.
In area (ii), no significant circular polarization signals are detected in the Mg~{\sc ii} $h$ and $k$ lines, and
many pixels are marked with black segments in Figure~\ref{fig:av_prof_rg2}.
This suggests that the magnetic field in the upper chromosphere would be nearly horizontal in this region.

\section{Conclusions and Discussions}
In this study, we used the suborbital rocket experiment CLASP2.1 observations to investigate the structure of the magnetic fields from the photosphere to the upper chromosphere in an active region.
CLASP2.1 obtained a 2D map of the Stokes spectra over 
a small FOV ($29\arcsec\times196\arcsec$) of
the observed active region with a variety of magnetic field structures such as a plage, a pore, and a penumbra, and we applied the WFA to infer the $B_{L}$ throughout the chromosphere.
The derived $B_{L}$ map is consistent with the one obtained by applying the HanleRT-TIC \citep{2024ApJ...974..154L}.
The $B_{L}$ mapping, combined with the photospheric magnetic field measurements conducted by the Hinode/SOT, 
allows us to quantitatively evaluate how the magnetic field changes with height  from the photosphere up to the upper chromosphere
and to directly investigate the causal relationship between the magnetic fields and the plasma structures in the upper atmospheric layers of the transition region and the corona. 

In the CLASP2.1 FOV we find a bright plage region dominated by negative polarity magnetic fields, where the so-called moss is observed in the transition region and the low-temperature corona  (e.g., AIA 171~{\AA} channel).
The magnetized area expands by a factor of $\sim2.5$ and $\sim3.1$ in the lower and middle chromosphere, respectively, compared to the photosphere. 
In the CLASP2.1 observation, the determination of the magnetized area in the upper chromosphere 
is difficult due to the noise. Nevertheless, the CLASP2 observation 
of an active region plage with its fixed slit position but with much longer exposure time (i.e.,
better S/N) shows that the spatial variation is similar between the middle and upper chromosphere \citep{2021SciA....7.8406I}, and it is conceivable that  
the magnetized area in the upper chromosphere is comparable to that of the middle chromosphere in CLASP2.1 as well.
Moreover, the $B_{L}$ in the middle chromosphere shows a high correlation with the moss intensity observed in the $k$ line core and the AIA~171~{\AA} images.
This suggests that the expanded magnetic fields at chromospheric heights 
could correspond to the moss structure in the transition region and the corona.
The moss region observed at EUV wavelengths are bright and dynamic areas around the footpoints of hot coronal loops in
active regions, and the chromospheric magnetic fields play a critical role in shaping the transition region and coronal plasma.

CLASP2.1 revealed a polarity reversal at the upper chromosphere within a positive-polarity pore embedded in a plage dominated by negative polarity.
The positive polarity region becomes larger in the lower and middle chromosphere compared to the photosphere. However, in the upper chromosphere, it becomes smaller, and a negative polarity area appears at the edge instead.
This localized polarity reversal, limited to the upper chromosphere, 
is indicative of an overlying background magnetic field (Figure~\ref{fig:cartoon}).
These magnetic configurations can induce magnetic discontinuities above the top chromosphere, possibly
leading to magnetic reconnection.
Indeed, as observed in SDO/AIA, the region around the positive magnetic patch is highly dynamic with recurrent transient brightenings and eruptions. 
The magnetic field in the upper chromosphere, revealed by the cores of the Mg~{\sc ii} $h$ and $k$ lines, provides critical information about the transition region and the corona.

The magnetic fields of chromospheric superpenumbral fibrils have been investigated with the Ca~{\sc ii} line at 8542~{\AA} \citep{2011A&A...527L...8D} and the He~{\sc i} triplet at 10830~{\AA} lines \citep{2013ApJ...768..111S}.
CLASP2.1 revealed the magnetic structure of the superpenumbral fibrils in the upper chromosphere with the Mg~{\sc ii} $h$ and $k$ lines.
The superpenumbral fibrils correspond to large-scale magnetic fields originating from the sunspot, 
existing only at the upper chromosphere,
as shown by the dominance of the positive $B_{L}$ fields at this height (Figure~\ref{fig:cartoon_sunspot}).
\cite{Song_CLASP21} found recurring loop brightening above the superpenumbral fibrils which seem to be associated with a magnetic discontinuity at the upper chromosphere.
In the lower chromosphere and below, the low-lying fields are routed in the sunspot penumbra and the nearby magnetic concentrations have the opposite polarity to the photospheric magnetic field in the sunspot.
The connectivity of the outer endpoints of the fibrils are not well established \citep{2013ApJ...768..111S}.
The CLASP2.1 FOV covers an area $<25\arcsec$ from the edge of the penumbra, 
and clear outer endpoints (negative polarity fields at the top chromosphere) 
are not found inside the CLASP2.1 FOV.
The endpoint would be located outside of the FOVs, most of the magnetic fields would reach the higher layers 
without returning to the photosphere, or the magnetic fields are so shallow that the outer endpoints are not detectable.
The absence of correlation between the intensity and the $B_{L}$ in the middle and upper chromosphere suggests that $B_{L}$ alone does not determine the structure of the superpenumbral fibrils.  
It is a reasonable conclusion, as the magnetic fields in the superpenumbral fibril are observed to be highly inclined \citep[e.g.,][]{2013ApJ...768..111S}, a configuration that is also supported by this study, as discussed below.

We studied also the possible signature of the impact of the Zeeman effect on the linear polarization
in the core of the Mg~{\sc ii} $h$ line by spatially averaging the profiles,
and the transverse magnetic fields at the upper chromosphere were estimated to be 
$1500$~G above the penumbra and $1000$~G along a superpenumbral fibril.
In these region, the longitudinal magnetic field strengths are significantly smaller than the transverse components, indicating that the magnetic fields are nearly horizontal in the upper chromosphere.
In observations with the Ca~{\sc ii} at 8542~{\AA} and He~{\sc i} at 10830~{\AA} lines,
field strengths larger than 1000~G have been inferred above the penumbra \citep{2016A&A...596A...8J},
whereas $B_{T}<350$~G are measured in the superpenumbral fibril regions \citep{2021A&A...649A.106Y}.
The large differences between the field strengths in the superpenumbral fibrils found in such works may be due to the difference in the regions that were observed, or due to the different instruments 
(i.e., spatial and spectral resolutions, SNR, and spectral lines).
Note that the estimated $B_{T}$ from the $Q$ and $U$ signals in the $h$ core has to be considered as an upper limit, 
because it is derived assuming that the 
observed linear polarization signals are fully caused by the Zeeman effect.
Further investigations considering the scattering of anisotropic radiation as well as the Zeeman effect would be required. 

Our estimations based on the WFA indicate that  the emergent linear polarization signals induced by the transverse magnetic field are very small 
($\sim0.1\%$ with $B_{T}\sim1500$~G) and it is difficult to detect signals outside of the sunspot region
where the field strength is expected to be less than 1000~G, 
even though a high S/N is achieved.
Such difficulty is due to the intrinsically shallow and wide spectral lines of Mg~{\sc ii} \citep{2020ApJ...901...32J}
as the linear polarization induced by the Zeeman effect is more strongly affected by the Doppler broadening than the circular polarization.
Complementary observations with other chromospheric lines, such as Ca~{\sc ii} at 8542~{\AA} and He~{\sc i} at 10830~{\AA}, whose linear polarization signals have some contribution from the Zeeman effect, would also be helpful.  

\begin{acknowledgments}
{\it Acknowledgements.} 
CLASP2.1 is an international partnership between NASA/MSFC, NAOJ, JAXA, IAC, and IAS; additional partners include ASCR, IRSOL, LMSAL, and the University of Oslo.
The Japanese participation was funded by JAXA as a Small Mission-of-Opportunity Program, JSPS KAKENHI (Grant numbers JP25220703, JP16H03963, JP19K03935, and JP21K01138), 2015 ISAS Grant for Promoting International Mission Collaboration, and by 2016 NAOJ Grant for Development Collaboration. The USA participation was funded by NASA Award $20-\mathrm{HLCAS}20-0010$. 
The Spanish participation was funded by the European  Research Council (ERC) through Advanced grand agreement No. 742265. 
The French hardware participation was funded by CNES funds CLASP2-13616A and 13617A.
L.B., J.T.B., and J.\v{S}. acknowledge support from the Swiss National
Science Foundation (SNSF) through grant CRSII5\_180238.
T.P.A.’s participation is part of the Project RYC2021-034006-I,
funded by MICIN/AEI/10.13039/501100011033, and the European Union “NextGenerationEU”/RTRP. 
T.P.A. and J.T.B. acknowledge support from the Agencia Estatal de 
Investigaci\'{o}n del Ministerio de Ciencia, Innovaci\'{o} y Universidades (MCIU/AEI) under grant “Polarimetric Inference of Magnetic Fields” and the European Regional Development Fund (ERDF) with reference PID2022-136563NB-I00/10.13039/501100011033.
B.D.P. was supported by NASA contract NNG09FA40C  (IRIS).
J.\v{S}. acknowledges the financial support from project \mbox{RVO:67985815} of the Astronomical Institute of the Czech Academy of Sciences.
Hinode is a Japanese mission developed and launched by ISAS/JAXA, 
with NAOJ as domestic partner and NASA and STFC (UK) as international partners. 
It is operated by these agencies in cooperation with ESA and NSC (Norway).
IRIS is a NASA Small Explorer Mission developed and operated by LMSAL 
with mission operations executed at NASA Ames Research Center 
and major contributions to downlink communications funded by ESA and the Norwegian Space Centre.
\end{acknowledgments}

%

\vspace{5mm}





\appendix



\section{Definition of Magnetized Area}\label{AD_D}
\begin{figure}[ht]
\begin{center}
	\includegraphics[angle=90,width=\linewidth]{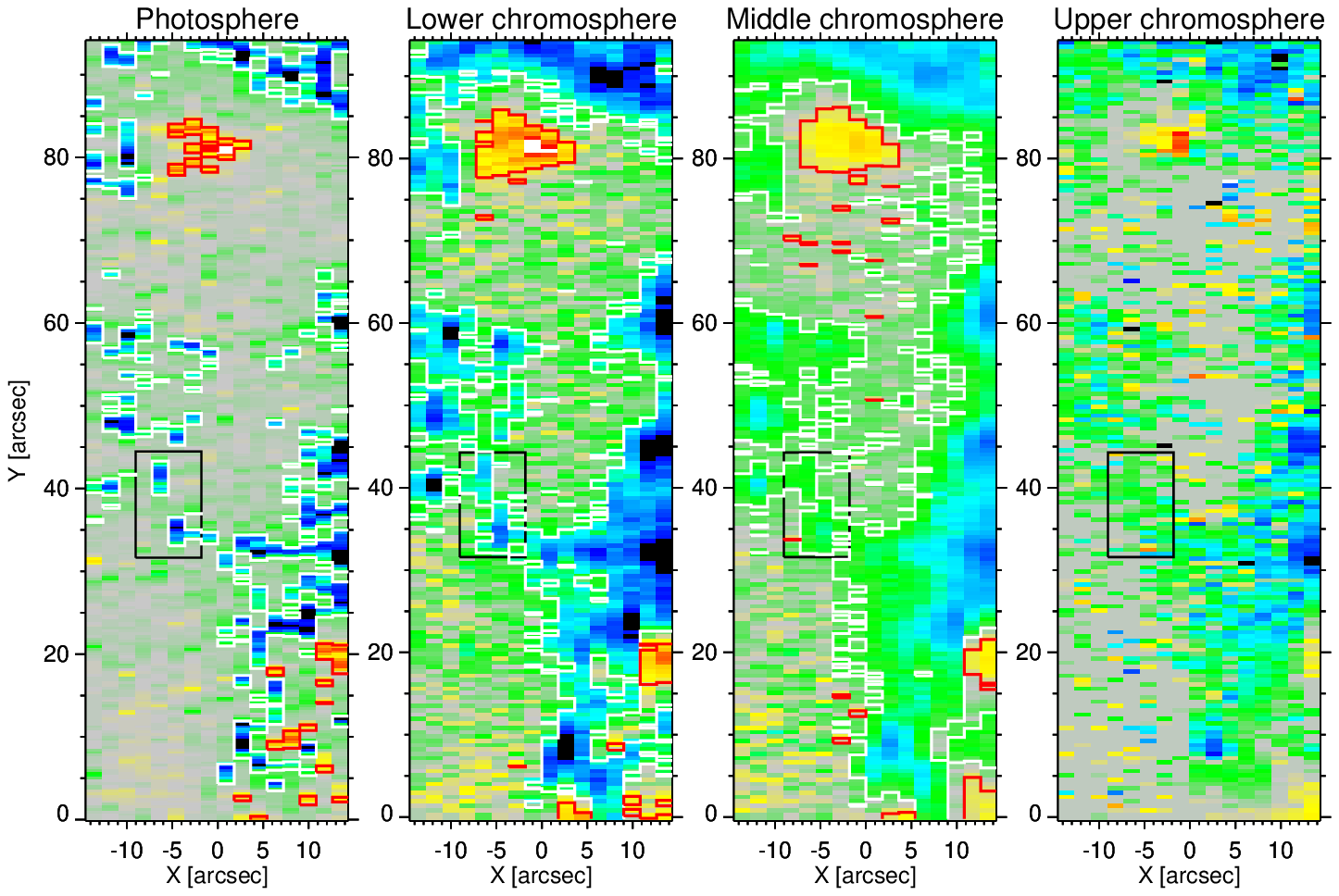}
\end{center}
	\caption{$B_{L}$ map focusing on the moss region
	($\mathrm{Y}>0\arcsec$ in Figure \ref{fig:blos}).
	The red and white contours show the pixels with 
	$B_{L}>0.2\max(B_{L})$ (positive polarity) and $B_{L}<0.2\min(B_{L})$ (negative polarity) that are defined 
	as the magnetized area. 
	The black box shows the area that is used to calculate the area ratio with the 		isolated magnetic structures. 
}
\label{fig:bl_area}
\end{figure}

Figure~\ref{fig:bl_area} shows the identified region (indicated by white contours) 
considered in the calculation of the magnetized area at each atmospheric height.

\section{Correlation Coefficients between $B_{L}$ and Intensity}\label{ap:cor}
Correlation coefficients between $B_{L}$ at four heights in the atmosphere and intensity with different wavelengths for the superpenumbral fibril region are shown in Table~\ref{tab:cc_f}.

\begin{table}[ht]
\centering
\begin{tabular}{c|ccccc}
 & \multicolumn{5}{c}{Correlation coefficient with intensity}\\
$B_{L}$ & 1600~{\AA} & $k$ core & 304~{\AA} & 171~{\AA} & 335~{\AA} \\
\hline
Upper chromosphere & -0.14 & 0.20 & 0.08 & 0.16 & 0.19\\
\hline
Middle chromosphere & 0.18 & 0.34 & 0.12 & 0.20 & 0.19\\
\hline
Lower chromosphere & 0.44 & 0.43 & 0.16 & 0.25 & 0.22\\
\hline
Photosphere & 0.63 & 0.22 & 0.03 & 0.09 & 0.06 \\
\hline
\end{tabular}
\caption{\label{tab:cc_f} Same as Table~\ref{tab:cc} but for the
superpenumbral fibril regions ($-50\arcsec\le\mathrm{Y}<0\arcsec$ in Figure~\ref{fig:blos}).}
\end{table}


\bibliography{ref}{}

\begin{thebibliography}{}
\expandafter\ifx\csname natexlab\endcsname\relax\def\natexlab#1{#1}\fi
\providecommand{\url}[1]{\href{#1}{#1}}
\providecommand{\dodoi}[1]{doi:~\href{http://doi.org/#1}{\nolinkurl{#1}}}
\providecommand{\doeprint}[1]{\href{http://ascl.net/#1}{\nolinkurl{http://ascl.net/#1}}}
\providecommand{\doarXiv}[1]{\href{https://arxiv.org/abs/#1}{\nolinkurl{https://arxiv.org/abs/#1}}}

\bibitem[{{Afonso Delgado} {et~al.}(2023){Afonso Delgado}, {del Pino
  Alem{\'a}n}, \& {Trujillo Bueno}}]{2023ApJ...954..218A}
{Afonso Delgado}, D., {del Pino Alem{\'a}n}, T., \& {Trujillo Bueno}, J. 2023,
  \apj, 954, 218, \dodoi{10.3847/1538-4357/ace4c8}

\bibitem[{{Alsina Ballester} {et~al.}(2016){Alsina Ballester}, {Belluzzi}, \&
  {Trujillo Bueno}}]{2016ApJ...831L..15A}
{Alsina Ballester}, E., {Belluzzi}, L., \& {Trujillo Bueno}, J. 2016, \apjl,
  831, L15, \dodoi{10.3847/2041-8205/831/2/L15}

\bibitem[{{Bate} {et~al.}(2024){Bate}, {Jess}, {Grant}, {Hillier}, {Skirvin},
  {Van Doorsselaere}, {Jafarzadeh}, {Wiegelmann}, {Duckenfield}, {Beck},
  {Moore}, {Stangalini}, {Keys}, \& {Christian}}]{2024ApJ...970...66B}
{Bate}, W., {Jess}, D.~B., {Grant}, S.~D.~T., {et~al.} 2024, \apj, 970, 66,
  \dodoi{10.3847/1538-4357/ad4d97}

\bibitem[{{Beck} \& {Choudhary}(2020)}]{2020ApJ...891..119B}
{Beck}, C., \& {Choudhary}, D.~P. 2020, \apj, 891, 119,
  \dodoi{10.3847/1538-4357/ab75bd}

\bibitem[{{Belluzzi} \& {Trujillo Bueno}(2012)}]{2012ApJ...750L..11B}
{Belluzzi}, L., \& {Trujillo Bueno}, J. 2012, \apjl, 750, L11,
  \dodoi{10.1088/2041-8205/750/1/L11}

\bibitem[{{Berger} {et~al.}(1999){Berger}, {De Pontieu}, {Fletcher},
  {Schrijver}, {Tarbell}, \& {Title}}]{1999SoPh..190..409B}
{Berger}, T.~E., {De Pontieu}, B., {Fletcher}, L., {et~al.} 1999, \solphys,
  190, 409, \dodoi{10.1023/A:1005286503963}

\bibitem[{{Borrero} \& {Ichimoto}(2011)}]{2011LRSP....8....4B}
{Borrero}, J.~M., \& {Ichimoto}, K. 2011, Living Reviews in Solar Physics, 8,
  4, \dodoi{10.12942/lrsp-2011-4}

\bibitem[{{Bose} {et~al.}(2024){Bose}, {De Pontieu}, {Hansteen}, {Sainz Dalda},
  {Savage}, \& {Winebarger}}]{2024NatAs...8..697B}
{Bose}, S., {De Pontieu}, B., {Hansteen}, V., {et~al.} 2024, Nature Astronomy,
  8, 697, \dodoi{10.1038/s41550-024-02241-8}

\bibitem[{{Campos Rozo} {et~al.}(2023){Campos Rozo}, {Vargas Dom{\'\i}nguez},
  {Utz}, {Veronig}, \& {Hanslmeier}}]{2023A&A...674A..91C}
{Campos Rozo}, J.~I., {Vargas Dom{\'\i}nguez}, S., {Utz}, D., {Veronig}, A.~M.,
  \& {Hanslmeier}, A. 2023, \aap, 674, A91, \dodoi{10.1051/0004-6361/202346389}

\bibitem[{{Carlsson} {et~al.}(2015){Carlsson}, {Leenaarts}, \& {De
  Pontieu}}]{2015ApJ...809L..30C}
{Carlsson}, M., {Leenaarts}, J., \& {De Pontieu}, B. 2015, \apjl, 809, L30,
  \dodoi{10.1088/2041-8205/809/2/L30}

\bibitem[{{Centeno} {et~al.}(2022){Centeno}, {Rempel}, {Casini}, \& {del Pino
  Alem{\'a}n}}]{2022ApJ...936..115C}
{Centeno}, R., {Rempel}, M., {Casini}, R., \& {del Pino Alem{\'a}n}, T. 2022,
  \apj, 936, 115, \dodoi{10.3847/1538-4357/ac886f}

\bibitem[{Chae {et~al.}(2014)Chae, Yang, Park, Maurya, Cho, \&
  Yurchysyn}]{Chae_2014}
Chae, J., Yang, H., Park, H., {et~al.} 2014, The Astrophysical Journal, 789,
  108, \dodoi{10.1088/0004-637X/789/2/108}

\bibitem[{{de la Cruz Rodr{\'\i}guez} \&
  {Socas-Navarro}(2011)}]{2011A&A...527L...8D}
{de la Cruz Rodr{\'\i}guez}, J., \& {Socas-Navarro}, H. 2011, \aap, 527, L8,
  \dodoi{10.1051/0004-6361/201016018}

\bibitem[{{De Pontieu} {et~al.}(1999){De Pontieu}, {Berger}, {Schrijver}, \&
  {Title}}]{1999SoPh..190..419D}
{De Pontieu}, B., {Berger}, T.~E., {Schrijver}, C.~J., \& {Title}, A.~M. 1999,
  \solphys, 190, 419, \dodoi{10.1023/A:1005220606223}

\bibitem[{{De Pontieu} {et~al.}(2014){De Pontieu}, {Title}, {Lemen}, {Kushner},
  {Akin}, {Allard}, {Berger}, {Boerner}, {Cheung}, {Chou}, {Drake}, {Duncan},
  {Freeland}, {Heyman}, {Hoffman}, {Hurlburt}, {Lindgren}, {Mathur}, {Rehse},
  {Sabolish}, {Seguin}, {Schrijver}, {Tarbell}, {W{\"u}lser}, {Wolfson},
  {Yanari}, {Mudge}, {Nguyen-Phuc}, {Timmons}, {van Bezooijen}, {Weingrod},
  {Brookner}, {Butcher}, {Dougherty}, {Eder}, {Knagenhjelm}, {Larsen},
  {Mansir}, {Phan}, {Boyle}, {Cheimets}, {DeLuca}, {Golub}, {Gates}, {Hertz},
  {McKillop}, {Park}, {Perry}, {Podgorski}, {Reeves}, {Saar}, {Testa}, {Tian},
  {Weber}, {Dunn}, {Eccles}, {Jaeggli}, {Kankelborg}, {Mashburn}, {Pust},
  {Springer}, {Carvalho}, {Kleint}, {Marmie}, {Mazmanian}, {Pereira}, {Sawyer},
  {Strong}, {Worden}, {Carlsson}, {Hansteen}, {Leenaarts}, {Wiesmann},
  {Aloise}, {Chu}, {Bush}, {Scherrer}, {Brekke}, {Martinez-Sykora}, {Lites},
  {McIntosh}, {Uitenbroek}, {Okamoto}, {Gummin}, {Auker}, {Jerram}, {Pool}, \&
  {Waltham}}]{2014SoPh..289.2733D}
{De Pontieu}, B., {Title}, A.~M., {Lemen}, J.~R., {et~al.} 2014, \solphys, 289,
  2733, \dodoi{10.1007/s11207-014-0485-y}

\bibitem[{{del Pino Alem{\'a}n} {et~al.}(2022){del Pino Alem{\'a}n}, {Alsina
  Ballester}, \& {Trujillo Bueno}}]{2022ApJ...940...78D}
{del Pino Alem{\'a}n}, T., {Alsina Ballester}, E., \& {Trujillo Bueno}, J.
  2022, \apj, 940, 78, \dodoi{10.3847/1538-4357/ac922c}

\bibitem[{{del Pino Alem{\'a}n} {et~al.}(2016){del Pino Alem{\'a}n}, {Casini},
  \& {Manso Sainz}}]{2016ApJ...830L..24D}
{del Pino Alem{\'a}n}, T., {Casini}, R., \& {Manso Sainz}, R. 2016, \apjl, 830,
  L24, \dodoi{10.3847/2041-8205/830/2/L24}

\bibitem[{{del Pino Alem{\'a}n} {et~al.}(2020){del Pino Alem{\'a}n}, {Trujillo
  Bueno}, {Casini}, \& {Manso Sainz}}]{2020ApJ...891...91D}
{del Pino Alem{\'a}n}, T., {Trujillo Bueno}, J., {Casini}, R., \& {Manso
  Sainz}, R. 2020, \apj, 891, 91, \dodoi{10.3847/1538-4357/ab6bc9}

\bibitem[{{Foukal}(1971)}]{1971SoPh...20..298F}
{Foukal}, P. 1971, \solphys, 20, 298, \dodoi{10.1007/BF00159759}

\bibitem[{{Foukal} {et~al.}(1974){Foukal}, {Noyes}, {Reeves}, {Schmahl},
  {Timothy}, {Vernazza}, {Wilhbroe}, \& {Huber}}]{1974ApJ...193L.143F}
{Foukal}, P.~V., {Noyes}, R.~W., {Reeves}, E.~M., {et~al.} 1974, \apjl, 193,
  L143, \dodoi{10.1086/181651}

\bibitem[{{Freeland} \& {Handy}(1998)}]{1998SoPh..182..497F}
{Freeland}, S.~L., \& {Handy}, B.~N. 1998, \solphys, 182, 497,
  \dodoi{10.1023/A:1005038224881}

\bibitem[{{Gordino} {et~al.}(2022){Gordino}, {Auch{\`e}re}, {Vial},
  {Bocchialini}, {Hassler}, {Bando}, {Ishikawa}, {Kano}, {Kobayashi},
  {Narukage}, {Trujillo Bueno}, \& {Winebarger}}]{2022A&A...657A..86G}
{Gordino}, M., {Auch{\`e}re}, F., {Vial}, J.~C., {et~al.} 2022, \aap, 657, A86,
  \dodoi{10.1051/0004-6361/202141960}

\bibitem[{{Hale}(1908)}]{1908ApJ....28..315H}
{Hale}, G.~E. 1908, \apj, 28, 315, \dodoi{10.1086/141602}

\bibitem[{Ichimoto {et~al.}(2008)Ichimoto, Lites, Elmore, Suematsu, Tsuneta,
  Katsukawa, Shimizu, Shine, Tarbell, Title, Kiyohara, Shinoda, Card, Lecinski,
  Streander, Nakagiri, Miyashita, Noguchi, Hoffmann, \& Cruz}]{Ichimoto2008}
Ichimoto, K., Lites, B., Elmore, D., {et~al.} 2008, Solar Physics, 249, 233

\bibitem[{{Ishikawa} {et~al.}(2013){Ishikawa}, {Kano}, {Bando}, {Suematsu},
  {Ishikawa}, {Kubo}, {Narukage}, {Hara}, {Tsuneta}, {Watanabe}, {Ichimoto},
  {Aoki}, \& {Miyagawa}}]{Ishikawa2013}
{Ishikawa}, R., {Kano}, R., {Bando}, T., {et~al.} 2013, \ao, 52, 8205,
  \dodoi{10.1364/AO.52.008205}

\bibitem[{{Ishikawa} {et~al.}(2021){Ishikawa}, {Bueno}, {del Pino Alem{\'a}n},
  {Okamoto}, {McKenzie}, {Auch{\`e}re}, {Kano}, {Song}, {Yoshida}, {Rachmeler},
  {Kobayashi}, {Hara}, {Kubo}, {Narukage}, {Sakao}, {Shimizu}, {Suematsu},
  {Bethge}, {De Pontieu}, {Dalda}, {Vigil}, {Winebarger}, {Ballester},
  {Belluzzi}, {{\v{S}}t{\v{e}}p{\'a}n}, {Ramos}, {Carlsson}, \&
  {Leenaarts}}]{2021SciA....7.8406I}
{Ishikawa}, R., {Bueno}, J.~T., {del Pino Alem{\'a}n}, T., {et~al.} 2021,
  Science Advances, 7, eabe8406, \dodoi{10.1126/sciadv.abe8406}

\bibitem[{{Ishikawa} {et~al.}(2023){Ishikawa}, {Trujillo Bueno}, {Alsina
  Ballester}, {Belluzzi}, {del Pino Alem{\'a}n}, {McKenzie}, {Auch{\`e}re},
  {Kobayashi}, {Okamoto}, {Rachmeler}, \& {Song}}]{2023ApJ...945..125I}
{Ishikawa}, R., {Trujillo Bueno}, J., {Alsina Ballester}, E., {et~al.} 2023,
  \apj, 945, 125, \dodoi{10.3847/1538-4357/acb64e}

\bibitem[{{Ishikawa} {et~al.}(2015){Ishikawa}, {Shimizu}, {Kano}, {Bando},
  {Ishikawa}, {Giono}, {Tsuneta}, {Nakayama}, \&
  {Tajima}}]{2015SoPh..290.3081I}
{Ishikawa}, S., {Shimizu}, T., {Kano}, R., {et~al.} 2015, \solphys, 290, 3081,
  \dodoi{10.1007/s11207-015-0774-0}

\bibitem[{{Jiang} {et~al.}(2012){Jiang}, {Fang}, \&
  {Chen}}]{2012ApJ...751..152J}
{Jiang}, R.~L., {Fang}, C., \& {Chen}, P.~F. 2012, \apj, 751, 152,
  \dodoi{10.1088/0004-637X/751/2/152}

\bibitem[{{Joshi} {et~al.}(2016){Joshi}, {Lagg}, {Solanki}, {Feller},
  {Collados}, {Orozco Su{\'a}rez}, {Schlichenmaier}, {Franz}, {Balthasar},
  {Denker}, {Berkefeld}, {Hofmann}, {Kiess}, {Nicklas}, {Pastor Yabar},
  {Rezaei}, {Schmidt}, {Schmidt}, {Sobotka}, {Soltau}, {Staude}, {Strassmeier},
  {Volkmer}, {von der L{\"u}he}, \& {Waldmann}}]{2016A&A...596A...8J}
{Joshi}, J., {Lagg}, A., {Solanki}, S.~K., {et~al.} 2016, \aap, 596, A8,
  \dodoi{10.1051/0004-6361/201629214}

\bibitem[{{Judge} {et~al.}(2020){Judge}, {Kleint}, {Leenaarts}, {Sukhorukov},
  \& {Vial}}]{2020ApJ...901...32J}
{Judge}, P.~G., {Kleint}, L., {Leenaarts}, J., {Sukhorukov}, A.~V., \& {Vial},
  J.-C. 2020, \apj, 901, 32, \dodoi{10.3847/1538-4357/abadf4}

\bibitem[{{Kano} {et~al.}(2017){Kano}, {Trujillo Bueno}, {Winebarger},
  {Auch{\`e}re}, {Narukage}, {Ishikawa}, {Kobayashi}, {Bando}, {Katsukawa},
  {Kubo}, {Ishikawa}, {Giono}, {Hara}, {Suematsu}, {Shimizu}, {Sakao},
  {Tsuneta}, {Ichimoto}, {Goto}, {Belluzzi}, {{\v{S}}t{\v{e}}p{\'a}n}, {Asensio
  Ramos}, {Manso Sainz}, {Champey}, {Cirtain}, {De Pontieu}, {Casini}, \&
  {Carlsson}}]{2017ApJ...839L..10K}
{Kano}, R., {Trujillo Bueno}, J., {Winebarger}, A., {et~al.} 2017, \apjl, 839,
  L10, \dodoi{10.3847/2041-8213/aa697f}

\bibitem[{{Katsukawa} \& {Tsuneta}(2005)}]{2005ApJ...621..498K}
{Katsukawa}, Y., \& {Tsuneta}, S. 2005, \apj, 621, 498, \dodoi{10.1086/427488}

\bibitem[{Kosugi {et~al.}(2007)Kosugi, Matsuzaki, Sakao, Shimizu, Sone,
  Tachikawa, Hashimoto, Minesugi, Ohnishi, Yamada, Tsuneta, Hara, Ichimoto,
  Suematsu, Shimojo, Watanabe, Shimada, Davis, Hill, Owens, Title, Culhane,
  Harra, Doschek, \& Golub}]{Kosugi2007}
Kosugi, T., Matsuzaki, K., Sakao, T., {et~al.} 2007, \solphys, 243, 3

\bibitem[{{Kubo} {et~al.}(2016){Kubo}, {Katsukawa}, {Suematsu}, {Kano},
  {Bando}, {Narukage}, {Ishikawa}, {Hara}, {Giono}, {Tsuneta}, {Ishikawa},
  {Shimizu}, {Sakao}, {Winebarger}, {Kobayashi}, {Cirtain}, {Champey},
  {Auch{\`e}re}, {Trujillo Bueno}, {Asensio Ramos}, {{\v S}t{\v e}p{\'a}n},
  {Belluzzi}, {Manso Sainz}, {De Pontieu}, {Ichimoto}, {Carlsson}, {Casini}, \&
  {Goto}}]{Kubo2016}
{Kubo}, M., {Katsukawa}, Y., {Suematsu}, Y., {et~al.} 2016, \apj, 832, 141,
  \dodoi{10.3847/0004-637X/832/2/141}

\bibitem[{{Landi Degl'Innocenti} \& {Landolfi}(2004)}]{2004ASSL..307.....L}
{Landi Degl'Innocenti}, E., \& {Landolfi}, M. 2004, {Polarization in Spectral
  Lines}, Vol. 307, \dodoi{10.1007/978-1-4020-2415-3}

\bibitem[{{Lemen} {et~al.}(2012){Lemen}, {Title}, {Akin}, {Boerner}, {Chou},
  {Drake}, {Duncan}, {Edwards}, {Friedlaender}, {Heyman}, {Hurlburt}, {Katz},
  {Kushner}, {Levay}, {Lindgren}, {Mathur}, {McFeaters}, {Mitchell}, {Rehse},
  {Schrijver}, {Springer}, {Stern}, {Tarbell}, {Wuelser}, {Wolfson}, {Yanari},
  {Bookbinder}, {Cheimets}, {Caldwell}, {Deluca}, {Gates}, {Golub}, {Park},
  {Podgorski}, {Bush}, {Scherrer}, {Gummin}, {Smith}, {Auker}, {Jerram},
  {Pool}, {Soufli}, {Windt}, {Beardsley}, {Clapp}, {Lang}, \&
  {Waltham}}]{2012SoPh..275...17L}
{Lemen}, J.~R., {Title}, A.~M., {Akin}, D.~J., {et~al.} 2012, \solphys, 275,
  17, \dodoi{10.1007/s11207-011-9776-8}

\bibitem[{{Li} {et~al.}(2024{\natexlab{a}}){Li}, {del Pino Alem{\'a}n}, \&
  {Trujillo Bueno}}]{2024ApJ...975..110L}
{Li}, H., {del Pino Alem{\'a}n}, T., \& {Trujillo Bueno}, J.
  2024{\natexlab{a}}, \apj, 975, 110, \dodoi{10.3847/1538-4357/ad7954}

\bibitem[{{Li} {et~al.}(2022){Li}, {del Pino Alem{\'a}n}, {Trujillo Bueno}, \&
  {Casini}}]{2022ApJ...933..145L}
{Li}, H., {del Pino Alem{\'a}n}, T., {Trujillo Bueno}, J., \& {Casini}, R.
  2022, \apj, 933, 145, \dodoi{10.3847/1538-4357/ac745c}

\bibitem[{{Li} {et~al.}(2023){Li}, {del Pino Alem{\'a}n}, {Trujillo Bueno},
  {Ishikawa}, {Alsina Ballester}, {McKenzie}, {Auch{\`e}re}, {Kobayashi},
  {Okamoto}, {Rachmeler}, \& {Song}}]{2023ApJ...945..144L}
{Li}, H., {del Pino Alem{\'a}n}, T., {Trujillo Bueno}, J., {et~al.} 2023, \apj,
  945, 144, \dodoi{10.3847/1538-4357/acb76e}

\bibitem[{{Li} {et~al.}(2024{\natexlab{b}}){Li}, {del Pino Alem{\'a}n},
  {Trujillo Bueno}, {Ishikawa}, {Alsina Ballester}, {McKenzie}, {Belluzzi},
  {Song}, {Okamoto}, {Kobayashi}, {Rachmeler}, {Bethge}, \&
  {Auch{\`e}re}}]{2024ApJ...974..154L}
---. 2024{\natexlab{b}}, \apj, 974, 154, \dodoi{10.3847/1538-4357/ad6dfb}

\bibitem[{Lites {et~al.}(2013)Lites, Akin, Card, Cruz, Duncan, Edwards, Elmore,
  Hoffmann, Katsukawa, Katz, Kubo, Ichimoto, Shimizu, Shine, Streander,
  Suematsu, Tarbell, Title, \& Tsuneta}]{Lites2013}
Lites, B.~W., Akin, D.~L., Card, G., {et~al.} 2013, \solphys, 283, 579

\bibitem[{{Loughhead}(1968)}]{1968SoPh....5..489L}
{Loughhead}, R.~E. 1968, \solphys, 5, 489, \dodoi{10.1007/BF00147015}

\bibitem[{{Madjarska}(2019)}]{2019LRSP...16....2M}
{Madjarska}, M.~S. 2019, Living Reviews in Solar Physics, 16, 2,
  \dodoi{10.1007/s41116-019-0018-8}

\bibitem[{{Mart{\'\i}nez Pillet} {et~al.}(1997){Mart{\'\i}nez Pillet}, {Lites},
  \& {Skumanich}}]{1997ApJ...474..810M}
{Mart{\'\i}nez Pillet}, V., {Lites}, B.~W., \& {Skumanich}, A. 1997, \apj, 474,
  810, \dodoi{10.1086/303478}

\bibitem[{{Morton} {et~al.}(2021){Morton}, {Mooroogen}, \&
  {Henriques}}]{2021RSPTA.37900183M}
{Morton}, R.~J., {Mooroogen}, K., \& {Henriques}, V.~M.~J. 2021, Philosophical
  Transactions of the Royal Society of London Series A, 379, 20200183,
  \dodoi{10.1098/rsta.2020.0183}

\bibitem[{{Nakamura} {et~al.}(2012){Nakamura}, {Shibata}, \&
  {Isobe}}]{2012ApJ...761...87N}
{Nakamura}, N., {Shibata}, K., \& {Isobe}, H. 2012, \apj, 761, 87,
  \dodoi{10.1088/0004-637X/761/2/87}

\bibitem[{{Pesnell} {et~al.}(2012){Pesnell}, {Thompson}, \&
  {Chamberlin}}]{2012SoPh..275....3P}
{Pesnell}, W.~D., {Thompson}, B.~J., \& {Chamberlin}, P.~C. 2012, \solphys,
  275, 3, \dodoi{10.1007/s11207-011-9841-3}

\bibitem[{{Rachmeler} {et~al.}(2022){Rachmeler}, {Trujillo Bueno}, {McKenzie},
  {Ishikawa}, {Auch{\`e}re}, {Kobayashi}, {Kano}, {Okamoto}, {Bethge}, {Song},
  {Alsina Ballester}, {Belluzzi}, {del Pino Alem{\'a}n}, {Asensio Ramos},
  {Yoshida}, {Shimizu}, {Winebarger}, {Kobelski}, {Vigil}, {De Pontieu},
  {Narukage}, {Kubo}, {Sakao}, {Hara}, {Suematsu}, {{\v{S}}t{\v{e}}p{\'a}n},
  {Carlsson}, \& {Leenaarts}}]{2022ApJ...936...67R}
{Rachmeler}, L.~A., {Trujillo Bueno}, J., {McKenzie}, D.~E., {et~al.} 2022,
  \apj, 936, 67, \dodoi{10.3847/1538-4357/ac83b8}

\bibitem[{{Schad} {et~al.}(2013){Schad}, {Penn}, \&
  {Lin}}]{2013ApJ...768..111S}
{Schad}, T.~A., {Penn}, M.~J., \& {Lin}, H. 2013, \apj, 768, 111,
  \dodoi{10.1088/0004-637X/768/2/111}

\bibitem[{{Scherrer} {et~al.}(2012){Scherrer}, {Schou}, {Bush}, {Kosovichev},
  {Bogart}, {Hoeksema}, {Liu}, {Duvall}, {Zhao}, {Title}, {Schrijver},
  {Tarbell}, \& {Tomczyk}}]{2012SoPh..275..207S}
{Scherrer}, P.~H., {Schou}, J., {Bush}, R.~I., {et~al.} 2012, \solphys, 275,
  207, \dodoi{10.1007/s11207-011-9834-2}

\bibitem[{{Schrijver} {et~al.}(1999){Schrijver}, {Title}, {Berger}, {Fletcher},
  {Hurlburt}, {Nightingale}, {Shine}, {Tarbell}, {Wolfson}, {Golub},
  {Bookbinder}, {DeLuca}, {McMullen}, {Warren}, {Kankelborg}, {Handy}, \& {De
  Pontieu}}]{1999SoPh..187..261S}
{Schrijver}, C.~J., {Title}, A.~M., {Berger}, T.~E., {et~al.} 1999, \solphys,
  187, 261, \dodoi{10.1023/A:1005194519642}

\bibitem[{Shimizu {et~al.}(2008)Shimizu, Nagata, Tsuneta, Tarbell, Edwards,
  Shine, Hoffmann, Thomas, Sour, Rehse, Ito, Kashiwagi, Tabata, Kodeki, Nagase,
  Matsuzaki, Kobayashi, Ichimoto, \& Suematsu}]{Shimizu2008}
Shimizu, T., Nagata, S., Tsuneta, S., {et~al.} 2008, \solphys, 249, 221

\bibitem[{{Shine} \& {Linsky}(1974)}]{1974SoPh...39...49S}
{Shine}, R.~A., \& {Linsky}, J.~L. 1974, \solphys, 39, 49,
  \dodoi{10.1007/BF00154970}

\bibitem[{{Sobotka} {et~al.}(2013){Sobotka}, {{\v{S}}vanda},
  {Jur{\v{c}}{\'a}k}, {Heinzel}, {Del Moro}, \&
  {Berrilli}}]{2013A&A...560A..84S}
{Sobotka}, M., {{\v{S}}vanda}, M., {Jur{\v{c}}{\'a}k}, J., {et~al.} 2013, \aap,
  560, A84, \dodoi{10.1051/0004-6361/201322148}

\bibitem[{{Song} {et~al.}(2018){Song}, {Ishikawa}, {Kano}, {Yoshida},
  {Tsuzuki}, {Uraguchi}, {Shinoda}, {Hara}, {Okamoto}, {Auch{\`e}re},
  {McKenzie}, {Rachmeler}, \& {Trujillo Bueno}}]{2018SPIE10699E..2WS}
{Song}, D., {Ishikawa}, R., {Kano}, R., {et~al.} 2018, in Society of
  Photo-Optical Instrumentation Engineers (SPIE) Conference Series, Vol. 10699,
  Space Telescopes and Instrumentation 2018: Ultraviolet to Gamma Ray, ed.
  J.-W.~A. {den Herder}, S.~{Nikzad}, \& K.~{Nakazawa}, 106992W,
  \dodoi{10.1117/12.2313056}

\bibitem[{{Song} {et~al.}(2022){Song}, {Ishikawa}, {Kano}, {McKenzie},
  {Trujillo Bueno}, {Auch{\`e}re}, {Rachmeler}, {Okamoto}, {Yoshida},
  {Kobayashi}, {Bethge}, {Hara}, {Shinoda}, {Shimizu}, {Suematsu}, {De
  Pontieu}, {Winebarger}, {Narukage}, {Kubo}, {Sakao}, {Asensio Ramos},
  {Belluzzi}, {{¥v{S}}t{\v{e}}p{\'a}n}, {Carlsson}, {del Pino Alem{\'a}n},
  {Alsina Ballester}, {Vigil}, \& {Leenaarts}}]{2022SoPh..297..135S}
{Song}, D., {Ishikawa}, R., {Kano}, R., {et~al.} 2022, \solphys, 297, 135,
  \dodoi{10.1007/s11207-022-02064-8}

\bibitem[{{Song} {et~al.}(2025){Song}, {Ishikawa}, {Kano}, {McKenzie},
  {Trujillo Bueno}, {Auch{\`e}re}, {Rachmeler}, {Okamoto}, {Yoshida},
  {Kobayashi}, {Bethge}, {Hara}, {Shinoda}, {Shimizu}, {Suematsu}, {De
  Pontieu}, {Winebarger}, {Narukage}, {Kubo}, {Sakao}, {Asensio Ramos},
  {Belluzzi}, {{\v{S}}t{\v{e}}p{\'a}n}, {Carlsson}, {del Pino Alem{\'a}n},
  {Alsina Ballester}, {Vigil}, \& {Leenaarts}}]{Song_CLASP21}
---. 2025, \apj, 978, 140, \dodoi{10.3847/1538-4357/ad94f6}

\bibitem[{Suematsu {et~al.}(2008)Suematsu, Tsuneta, Ichimoto, Shimizu, Otsubo,
  Katsukawa, Nakagiri, Noguchi, Tamura, Kato, Hara, Kubo, Mikami, Saito,
  Matsushita, Kawaguchi, Nakaoji, Nagae, Shimada, Takeyama, \&
  Yamamuro}]{Suematsu2008}
Suematsu, Y., Tsuneta, S., Ichimoto, K., {et~al.} 2008, \solphys, 249, 197

\bibitem[{{Testa} {et~al.}(2013){Testa}, {De Pontieu}, {Mart{\'\i}nez-Sykora},
  {DeLuca}, {Hansteen}, {Cirtain}, {Winebarger}, {Golub}, {Kobayashi},
  {Korreck}, {Kuzin}, {Walsh}, {DeForest}, {Title}, \&
  {Weber}}]{2013ApJ...770L...1T}
{Testa}, P., {De Pontieu}, B., {Mart{\'\i}nez-Sykora}, J., {et~al.} 2013,
  \apjl, 770, L1, \dodoi{10.1088/2041-8205/770/1/L1}

\bibitem[{{Tian} {et~al.}(2018){Tian}, {Samanta}, \&
  {Zhang}}]{2018GSL.....5....4T}
{Tian}, H., {Samanta}, T., \& {Zhang}, J. 2018, Geoscience Letters, 5, 4,
  \dodoi{10.1186/s40562-018-0103-1}

\bibitem[{Tsuneta {et~al.}(2008)Tsuneta, Ichimoto, Katsukawa, Nagata, Otsubo,
  Shimizu, Suematsu, Nakagiri, Noguchi, Tarbell, Title, Shine, Rosenberg,
  Hoffmann, Jurcevich, Kushner, Levay, Lites, Elmore, Matsushita, Kawaguchi,
  Saito, Mikami, Hill, \& Owens}]{Tsuneta2008}
Tsuneta, S., Ichimoto, K., Katsukawa, Y., {et~al.} 2008, \solphys, 249, 167

\bibitem[{{Tsuzuki} {et~al.}(2020){Tsuzuki}, {Ishikawa}, {Kano}, {Narukage},
  {Song}, {Yoshida}, {Uraguchi}, {Okamoto}, {McKenzie}, {Kobayashi},
  {Rachmeler}, {Auchere}, \& {Trujillo Bueno}}]{2020SPIE11444E..6WT}
{Tsuzuki}, T., {Ishikawa}, R., {Kano}, R., {et~al.} 2020, in Society of
  Photo-Optical Instrumentation Engineers (SPIE) Conference Series, Vol. 11444,
  Society of Photo-Optical Instrumentation Engineers (SPIE) Conference Series,
  114446W, \dodoi{10.1117/12.2562273}

\bibitem[{{Yadav} {et~al.}(2021){Yadav}, {D{\'\i}az Baso}, {de la Cruz
  Rodr{\'\i}guez}, {Calvo}, \& {Morosin}}]{2021A&A...649A.106Y}
{Yadav}, R., {D{\'\i}az Baso}, C.~J., {de la Cruz Rodr{\'\i}guez}, J., {Calvo},
  F., \& {Morosin}, R. 2021, \aap, 649, A106,
  \dodoi{10.1051/0004-6361/202039857}

\bibitem[{{Yoshida} {et~al.}(2018){Yoshida}, {Song}, {Ishikawa}, {Kano},
  {Katsukawa}, {Suematsu}, {Narukage}, {Kubo}, {Shinoda}, {Okamoto},
  {McKenzie}, {Rachmeler}, {Auch{\`e}re}, \& {Trujillo
  Bueno}}]{2018SPIE10699E..30Y}
{Yoshida}, M., {Song}, D., {Ishikawa}, R., {et~al.} 2018, in Society of
  Photo-Optical Instrumentation Engineers (SPIE) Conference Series, Vol. 10699,
  Space Telescopes and Instrumentation 2018: Ultraviolet to Gamma Ray, ed.
  J.-W.~A. {den Herder}, S.~{Nikzad}, \& K.~{Nakazawa}, 1069930,
  \dodoi{10.1117/12.2312463}

\end{thebibliography}
\bibliographystyle{aasjournal}



\end{document}